# IoT Data Discovery: Routing Table and Summarization Techniques

Hieu Tran, Son Nguyen, Student *Member, IEEE,* I-Ling Yen, Farokh Bastani *Fellow, IEEE*

**Abstract**— In this paper, we consider the IoT data discovery problem in very large and growing scale networks. Through analysis, examples, and experimental studies, we show the importance of peer-to-peer, unstructured routing for IoT data discovery and point out the space efficiency issue that has been overlooked in keyword based routing algorithms in unstructured networks. Specifically, as the first in the field, this paper investigates routing table designs and various compression techniques to support effective and space efficient IoT data discovery routing. Novel summarization algorithms, including alphabetical based, hash based, and meaning based summarization and their corresponding coding schemes, are proposed. We also consider routing table design to support summarization without degrading lookup efficiency for discovery query routing. The issue of potentially misleading routing due to summarization is also investigated. Subsequently, we analyze the strategy of when to summarize to balance the tradeoff between the routing table compression rate and the chance of causing misleading routing. For the experimental study, we have collected 100K IoT data streams from various IoT databases as the input dataset. Experimental results show that our summarization solution can reduce the routing table size by 20 to 30 folds with a 2-5% increase in latency compared with similar peer-to-peer discovery routing algorithms without summarization. Also, our approach outperforms DHT-based approaches by 2 to 6 folds in terms of latency and traffic.

―――――――――――――― ◆ ――――――――――――――

## 1 INTRODUCTION

The rapidly increasing number of IoT devices results in creating a torrent of data. These data may be stored and managed at the edge or regional sites of the Internet. For example, many smart cities may host their IoT data in municipal databases. Many manufacturers may host their IoT data at their sites. Data from smart vehicles will probably stay with the vehicles or roadside units. As can be seen, many IoT data are being created and hosted in a dispersed way over the Internet. These peer-to-peer (p2p) data sources can be discovered and used to achieve more advanced data analysis and knowledge discovery.

To enable IoT data stream discovery, annotation is an important issue. Discovery literature has considered raw keywords and multi-attribute annotation (MAA) models for annotation [1]. Many machine-interpretable semantic models for annotating the IoT data streams, such as the semantic sensor network, Fiesta-IoT, etc., have been proposed. However, complex annotations may significantly degrade the IoT data discovery performance. In this work, we consider MAA-based IoT data annotation and query specifications. The semantic based annotations can be converted into MAA to facilitate efficient routing.

Data discovery in p2p networks under keywords and MAA-based annotations has been widely investigated. Many existing works are DHT (distributed hash table) based [2], which hashes data objects to specific nodes in the network. They are very effective for some objects but are not suitable for IoT data. Basic DHT schemes require data to be stored at the hashed locations. IoT data streams are generally continuous flows, and moving them to potentially far away hosts based on DHT can cause major overhead. Some works use the hashed nodes as the pointers pointing to the actual IoT data locations [1] [3], but these approaches imply space and communication overhead, especially in handling MAA-based lookups.

Unstructured discovery routing approaches, such as [4] [5], require the maintenance of a routing table or routing information cache. These discovery routing schemes are similar to IP-based routing, except that the routing tables and the discovery queries are indexed by keywords instead of IP addresses. One crucial issue in unstructured p2p routing is the design of the routing table for maintaining or caching the routing information. Since resource-constrained nodes host many IoT data streams at the edge of the network, routing table size on these nodes must be confined. In IP networking, space concern is addressed by compressing the routing tables using the "summarization" technique [6]. However, very few works have considered space efficient routing table design or summarization techniques.

In this paper, we consider data discovery in p2p IoT networks at the edge of the Internet (including resource-constrained edge servers and IoT nodes). We focus on the space efficiency issue in query routing for discovery of data streams annotated by the MAA model. Specifically, we introduce the summarization techniques, and the corresponding routing table designs to achieve space efficient IoT data stream discovery in IoT-DBNs.

**How to summarize keywords used in data discovery routing**? Summarization in IP-based routing relies on the hierarchical structure of the domain names [6].

―――――――――

- *The authors are with the Department of Computer Science, the University of Texas at Dallas, Richardson, TX 75080.*
- *E-mail: {trunghieu.tran, sonnguyen, ilyen, Farokh.Bastani} @utdallas.edu.*

*Manuscript received May 2022;*

*(Corresponding author: Hieu Tran)*

Summarization of numerical values can also be done intuitively by union of individual ranges [5]. For summarizing keywords, a natural extension of these methods is *alphabetical based* summarization, which compresses multiple keywords into their longest common prefix. However, alphabetical based method depends on how likely we can find proper keyword groups that can be summarized into common prefixes. If the group of keywords that can be summarized does not show up together in a routing table, then it will be difficult to compress the table. In fact, experimental results show that the effectiveness of alphabetical based summarization is limited.

To avoid the dependency on keyword distributions, we can use hashing to randomize the set of keywords in the system to a uniformly distributed coding space to allow summarization to occur more uniformly. Accordingly, we also consider the *hash based* summarization in this paper.

When considering keyword distributions carefully, we can see that relevant keywords may have some regionalized appearance. For example, some cities are major manufacturing sites for certain products, such as Detroit for automotive manufacturing, Chicago for steel industry, etc. Specific sensor data streams for production line monitoring for particular industries may include many semantically similar keywords. In a smart city setting, sensors with similar functionalities, such as traffic observation, pollution monitoring, trash tracking, etc., may be used throughout the city. The descriptors for these sensor data streams will also have significant similarities. To achieve potentially better routing table compression by making use of the regionalized keyword similarity distributions, we also consider the *meaning based* summarization to summarize IoT data descriptors with similar keywords and expect that it will yield a better compression ratio.

**Is summarization for keywords feasible**? With keywords indexing the routing tables, how do we know which keywords (routing table entries) can be aggregated? This is straightforward for IP based routing, for numerical ranges, and for the alphabetical based method, but difficult in other approaches. Having keywords such as temperature, pressure, flow rate, etc., in a routing table, how do we know whether they can be summarized? No work has specifically addressed this issue.

We design novel coding schemes to maintain parent-child summarization relations to enable the identification of the child keywords that can be summarized and the parent keyword to be summarized into. These coding schemes can further improve space efficiency of the routing tables. Thus, we allow routing information to be summarized instead of thrown away to achieve space efficiency as well as routing effectiveness.

**How to know when to summarize**? Unlike IP based routing, where same domain prefixes have a high probability of implying the same routing directions, keyword based summarization does not have the same implication. Assume that a parent keyword $p$ has three child keywords $x$, $y$, and $z$ (i.e., $x$, $y$, and $z$ can be summarized into $p$). During summarization, how do we know $p$ has three child keywords? This is an issue even for alphabetical based summarization. Again, we need to maintain the ontology at each routing node, which is not feasible. Also, should we require that summarization to $p$ can only be done when all the 3 child keywords appear in the routing table (full coverage) or allow a partial set of these keywords to be summarized into $p$ (partial coverage). With IP addressing, this is less a problem due to its domain based design. But in keyword based routing, requiring full coverage reduces the routing table compression ratio and allowing partial coverage may result in misleading routing. In this paper, we not only analyze the impact of the "coverage" threshold, but also develop novel schemes to compute coverage for truly realizing the summarization strategies.

**Discovery routing table**. We develop novel data structures and algorithms for the routing tables to further enhance space efficiency in routing and facilitate efficient lookup for query routing and efficient table updates for routing information updates and summarization.

**Experimental study**. We crawled the web, obtained 100K IoT data streams, and extracted their descriptors [7]. Accordingly, we conducted in-depth experiments to study the performance of our summarization strategies under different settings. We also compared our approach with DHT-based and other unstructured routing solutions in large-scale IoT networks. Experimental results show that our summarization solutions yield great space efficiency for routing tables. A preliminary version of this paper appears in [8].

In the rest of the paper, Section 2 discusses the related literature. Section 3 formally defines our multi-attribute annotation model and our discovery routing problem. Section 4 discusses the three summarization strategies and the algorithmic designs to realize them. Section 5 considers the summarization coverage issue and discusses the schemes to enable coverage computation. Section 6 discusses handling the IoT-DBN growth. Section 7 introduces novel designs of the routing table and the routing algorithms to realize the proposed summarization techniques in a space and time-efficient way. Experimental results are presented in Section 8 and 9. Section 10 concludes the paper.

## 2 RELATED WORKS

Centralized IoT data discovery solutions [8], [9] require a tremendous amount of resources to maintain the information of all the IoT data sources worldwide. With the fast-growing number of IoT devices and their corresponding raw and processed data, the scalability may become a concern. More importantly, if the change rate of the IoT data streams is significant, it will take a lot higher communication cost to keep the centralized information up to date. Also, locality sometimes is important in the discovery process. Thus, peer-to-peer (p2p) discovery still plays an important role. There are two major directions in p2p data discovery, DHT based and unstructured routing, which were mainly designed for document search.

**DHT based solutions for MAA based data discovery**. A lot of works in keyword based discovery in peer-to-peer systems are based on distributed hash table (DHT) [2] [10]. However, in principle, DHT only handles discovery based on a single keyword. Several works extend DHT for handling object lookup with each object indexed by multiple keywords or MAA. Some works consider hashing individual keywords to the DHT. The querier has to retrieve all the matching object indices for each keyword in the query, determine the real match from the responses, and retrieve the object. [3] is such a solution. This approach incurs a heavy communication cost, i.e., $l$ round communication trips for a query with $l$ keywords. Also, a lot of information passed back may not be useful toward the query responses.

Other DHT-based works consider hashing all the powersets of the keyword set (or attribute-value pairs) for each object into the DHT. For example, each object may be specified by 10 attributed keywords, but the query may give any subset of the 10 keywords. Thus, $2^{10}-1$ keyword sets are hashed to the DHT to serve as pointers. [1] [11] are solutions based on this approach with various improvements.

[12] introduces the MKey scheme, which is a two-level approach. At the high level, a set of super-peers are structured by DHT. Each super-peer is associated to a cluster of peers and unstructured routing is used in the cluster. Same as other multi-keyword DHT schemes, object indices are still duplicated, and lookup needs to explore some DHT clusters that may not have the matching object.

Though DHT based approaches maintain exact information of where to forward a query, most of them still require $\log N$ overlay routing steps and each overlay step traverses half of the network (= $1/2\sqrt[k]{N}$ nodes, where $k$ is the connectivity degree) in average. For document search, DHT requires the objects being placed at their hashed locations, which is infeasible for IoT data discovery since IoT data streams are collected continuously. Moreover, basic DHT only handles a single keyword and MAA based approaches, as discussed above, further introduce space and communication overheads.

**Unstructured peer-to-peer networks and summarization techniques**. Information centric networking (ICN) consists of a set of keyword-based unstructured discovery routing schemes. In DONA [13], a public key based naming has been considered. It uses centralized or hierarchical Resolution Handler (RH) for naming resolution and IP like forwarding. In Content Centric Networking (CCN) [14] [15] and its successor Named Data Networking (NDN) [16], the object naming is by the URI like hierarchical descriptors and the routing protocols are similar to the conventional IP based approaches. These naming schemes are not suitable for the discovery of IoT data streams. Although hierarchical naming can merge different attributes into one string, it cannot handle missing attributes or allow attributes to be specified without a predetermined order. Also, none of the ICN works consider the feasible summarization techniques.

The GSD scheme [4] is an early unstructured peer-to-peer information caching solution and several similar works provide various improvements. None of these works consider summarization techniques or how to compress routing tables.

The Combined Broadcast and Content Based (CBCB) routing [5] is an unstructured peer-to-peer solution that can be used for object discovery. It considers a covering concept, i.e., if the newly advertised routing data is "covered" by existing routing data, then there will be no updates and the advertised information will not be forwarded further. Also, it merges multiple routing table indices if they can be expressed by an index covering all of them. Some claim this to be a form of summarization. But this intuitive technique can only be applied to numerical ranges effectively. When considering keywords, CBCB considers combining them by "and" and "or" operators, which may even increase the number of entries in the routing table. Thus, the space efficiency issue is not addressed.

Most of the unstructured p2p routing schemes adapt IP-based routing protocols by replacing IP addresses with keywords, but almost none of them consider the summarization techniques as their counterparts do because it is easier to summarize IP addresses, but difficult to summarize keywords. We use the AODV (ad hoc on-demand distance vector) IP-routing protocol with MAA but focusing on keyword-based summarization to reduce space overhead.

## 3 PROBLEM SPECIFICATION AND EXAMPLES

### 3.1 System Architecture

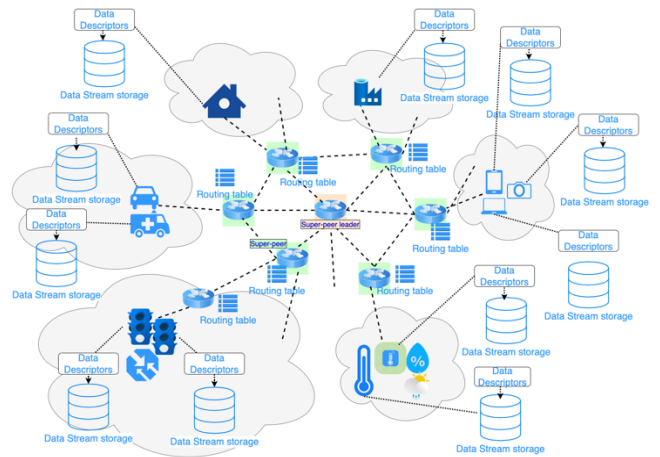

Fig. 1. *IoT database network architecture*

Fig. 1 shows the system architecture for the p2p IoT database network (IoT-DBN). IoT-DBN has a hierarchical structure. At the edge of the network are the IoT devices (sensor nodes) that observe some physical entities or the environments and generate continuous data streams, which are stored at the devices or the edge nodes. Each data stream is described by a set of data descriptors which serves as the metadata for data discovery. New data for an existing data stream will simply be stored with its metadata without changing the corresponding routing information. In case a large data stream is hosted by multiple nodes, its storage locations will be managed within its metadata. When a new data stream is generated, a new set of descriptors for the data stream (the metadata) needs to be advertised and the routing information in IoT-DBN needs to be updated.

The devices with similar characteristics (e.g., same locations, similar or correlated in what is observed, etc.) can be grouped into clusters and a super-peer can manage each cluster. The super-peers can be further clustered and the next layer super-peers can be used to manage the sub-clusters. Super-peer based p2p networks is a common structure [12] [18] and will not be discussed in detail. The specific Information to be maintained by the super-peers depends on the schemes and will be discussed later.

### 3.2 Example Cases

We consider data discovery in the p2p IoT database network (IoT-DBN), which consists of a huge number of individual IoT data streams and databases (each hosts multiple data streams) dispersed over the regional and edge of the Internet. A few example cases are given to illustrate IoT-DBN and the system model.

**Case 1**. Consider a scenario where an ambulance is requested for some injured victim of an accident. The goal is to get the ambulance that can arrive at the site at the earliest time. Various Map Apps may estimate driving time for regular cars, but not for ambulances. If the data is not directly available, we may collect ambulance driving data from other cities in similar traffics to derive the potential ambulance driving time.

**Case 2**. Consider that a police officer receives a report of a hit-and-run incident that occurred 5 minutes ago by a red sedan at a location with coordinate $(x, y)$. The police officer can request the recorded videos, if available, from vehicles that are within 5-minute driving range from $(x, y)$ to track the suspect.

**Case 3**. Performing data analytics based on multiple similar data sources can help significantly improve the analysis accuracy and help with cold start situations. For example, a combustion engine production company may retrieve sanitized data from noncompeting companies with similar production lines to help with its analytics for early fault detection and diagnosis of its production line. If the company is new and does not have its data, existing data can help avoid cold start.

**Example for the IoT-DBN model**. Consider Case 1 with sample GPS data for real-time traffic tracking provided by, for instance, the City of Cincinnati, the City of Columbus, or the City of Cleveland, etc. The annotation for a sample data stream is shown below.

(DataCategory: *GPS*; Vehicle type: *car*;
City: *Cincinnati*, Region: *Central Business District*;
Day: *Weekend*; Region traffic volume: *v*;
Duration: *3/2/18 17:03:20 - 3/2/18 19:10:30*)

The data streams may be hosted by individual cities and may still be in collection or may have finished collection. (Note: for Examples 2 and 3, the data are hosted by individual cars and individual companies, respectively.) Each IoT data stream is annotated, and the annotation (aka metadata) is a set of attribute-value pairs (descriptors). The example annotation contains static as well as dynamic attributes. The dynamic attributes are continuously updated as the new data flow in. "Region traffic volume" is an example of dynamic attributes, specified as the average number of vehicles per unit time passing through this region and computed up to the current time if the duration has not ended. Attribute "Vehicle type" is an example of static attributes and for Case 1, we need to consider types *car* and *ambulance*. (Note: Individual data objects in the data streams are annotated by GPS, timestamp, etc., but are not part of the data stream description.)

To enable IoT-DBN capabilities, we add a thin routing layer above individual distributed IoT data streams and/or databases to parse the data streams for attribute consistency and construct routing tables and perform discovery routing.

Following Case 1, a policeman from, e.g., Indianapolis, may issue queries to retrieve GPS data for vehicles of type *sedan* and *ambulance* of nearby cities with any period to derive the correlations and predict the driving time for ambulances in the accident region (which is assumed to be not available). The query could be described as following:

(DataCategory: *GPS*; Vehicle type: *ambulance* || *car*;
Region traffic volume: $[v_1, v_2]$)

The query requests data from any region and any duration with a traffic volume that is similar to the current condition. It may return Cincinnati's ambulance and car GPS data, which can help build the mapping between their driving times. The mapping can then be used to predict the ambulance driving time in Indianapolis assuming that only car GPS data is available in this city and the ambulance data is missing. We applied transfer learning techniques to make a similar prediction (for service vehicles instead of ambulance). It yields a high accuracy (MAPE=13.8) and takes only a few seconds to derive the mapping (based on 380K data points) and make the prediction. (The mapping is derived at the data source and only the model properties need to be transferred.)

### 3.3 Problem Specification

We consider the data discovery routing problem in IoT-DBN. For a data stream $ds$, let

$$AN^{ds} = \left((a_1: v_1^{ds}), (a_2: v_2^{ds}), \dots, (a_n: v_n^{ds})\right)$$

denote the annotation (key-value pairs) for $ds$, where $(a_i: v_i^{ds})$ is the "*descriptor*" for the $i$-th *attribute* $a_i$, and $v_i^{ds}$ is the *attribute value* for $a_i$ in $ds$'s description. Note that $AN^{ds}$ specifies a data stream $ds$ that plays a role as its *metadata* for discovery purpose, not individual data points. And, data discovery is based on the descriptions of data, i.e., $AN^{ds}$.

Let $AV_{a_i}^{sys}$, $1 \le i \le n$, denote the set of descriptors (attribute-value pairs) for attribute $a_i$ in $AN^{ds}$, for all $ds$ of all involved IoT databases in an IoT-DBN. $AV_{a_i}^{sys}$ may change dynamically, but it is relatively stable, and we consider $AV_{a_i}^{sys}$ at a specific time. We assume that an IoT-DBN is a specific application domain and there exists a common set of attributes. Even if some different attributes may appear, they are aligned, and the union of the attributes are considered. Also, it is possible that for some data streams, the values for some attributes are missing, i.e., $v_i^{ds} = \emptyset$.

Users may issue queries to look for data streams in the IoT-DBN. A query is specified by descriptors with a matching threshold. Specifically, $q = (A^q, \alpha, q_{src})$, where

$$A^q = \left((a_1: v_1^q), (a_2: v_2^q), \dots, (a_n: v_n^q)\right),$$

$\alpha$ is a threshold for query matching (will be defined later) and $q_{src}$ is the querier node. Also, $v_i^q = \phi$ if $a_i$ is not given in $q$.

**Attr-match**. Consider a descriptor in a query $q$, $(a_i: v_i^q)$, and a descriptor for a data stream $ds$, $(a_i: v_i^{ds})$. They have an attr-match for attribute $a_i$ iff $v_i^q = v_i^{ds} \vee v_i^q = \phi \vee v_i^{ds} = \phi$. Note that we can define operation "=" differently for different usage requirements.

In IoT-DBN, each database node $D$ maintains a routing table $RT^D$. $RT^D$ consists of a set of entries $((a: v), D')$, where $(a: v)$ is a descriptor and $D'$ is the forwarding neighbor through whom a node hosting a data stream $ds$ with matching $(a: v)$ can be reached. Note that the entries in $RT^D$ are organized in a certain data structure for efficient lookup (will be discussed in Sec. VI).

**$\alpha$-match**. Consider a query $q$ and a data stream $ds$ ($q$ and $ds$ are as defined earlier). We define $match_i = 1$ if $(a_i: v_i^q)$ and $(a_i: v_i^{ds})$ have an attr-match, and $match_i = 0$, otherwise. $ds$ has an $\alpha$-match with $q$ if and only if $\sum_{i=1}^n match_i \geq \alpha \times n$, $0 \leq \alpha \leq 1$. When $\alpha = 1$, each attribute provided in $q$ should have a match in $ds$. Similarly, consider an IoT-DBN node $D$. If $\exists D'$, such that $RT^D$ includes at least $\alpha \times n$ different entries $((a_i: v_i^q), D')$, for some $i$, then $D'$ is an $\alpha$-match forwarding neighbor of $D$ for $q$.

**Discovery routing**. Our goal is to perform discovery routing for query $q$ to discover data streams $ds$ by matching data stream annotations $AN^{ds}$. A routing mechanism consists of the advertisement protocol and the query forwarding protocol. We consider both table based routing and on-demand information caching (we call information cache also as routing table for consistency). Also, we consider the common routing protocol with our goals and assumptions.

*Advertisement protocol (AdvP)*. The advertisement phase is for routing table construction. Generally, it is performed when there are new data streams. During advertisement, node $D$ sends an advertisement message $advmsg$, which includes the descriptors of new data streams, to all its neighbors. When $D'$ receives $advmsg$ from $D$, for each descriptor $(a: v)$, an entry $((a: v), D)$ is added to $RT^{D'}$ if it does not already exist. The system gives a bound $b^{ad}$, which is the number of hops an advertisement message can traverse. If the hop count for $advmsg$ has not reached the bound $b^{ad}$, it will be forwarded to all $D'$'s neighbors except those listed as the forwarding neighbors for the descriptors in $advmsg$.

*Query Forwarding protocol (QFP)*. The system gives a bound on the hop counts for query forwarding, denoted by $b_q$. Consider a node $D$ receiving a query $q$ from $D'$. Let $R^{q,D} = (ds_1, ds_2, ..., ds_r)$ denote the data streams in $D$'s database that have $\alpha$-match with $q$. If $R^{q,D} \neq \phi$, then $D$ sends response $(R^{q,D}, q, D)$ back to $q_{src}$. In table driven routing, $R^{q,D}$ can be sent directly to $q_{src}$. In information caching based approaches, $R^{q,D}$ is delivered by a reverse route of query forwarding. Also, let $mns^{q,D} = \{D_1, ..., D_m\}$ denote the set of $\alpha$-matching neighbors for $q$ at node $D$. $D$ forwards $q$ to the neighbors in $mns^{q,D}$. If $\|mns^{q,D}\| = 0$ or the hop bound $b^q$ has been reached, then $D$ will not forward $q$. As can be seen, if there are no updates and $b^{ad} = \infty$ and $b_q = \infty$, then QFP will return all $\alpha$-matching data streams for $q$.

## 4 SUMMARIZATION TECHNIQUES

To enable data stream lookup in an IoT-DBN, a routing table $RT^D$ of node $D$ maintains $AN^{ds}$ for each data stream $ds$ whose advertisement reaches $D$. Since many nodes in the IoT-DBN may be resource-constrained, we consider space efficiency for routing tables. Specifically, we consider the summarization technique for routing table compression.

### 4.1 Common Attribute Based Compression

A basic approach for maintaining routing information for the data streams is to maintain all the attributes in $AN^{ds}$ for each $ds$. If we index the routing table by $AN^{ds}$ directly, then when routing a query $q$, local table matchmaking may be slow since we need to consider the potential of missing attributes in $q$ from $AN^{ds}$, for each $ds$. Also, there can be a huge number of routing table entries.

An alternate method is to index the routing table by individual descriptors in $AN^{ds}$. In this case, should we include the identity of $ds$ in the index? If so, each entry becomes $((a: v), ds, D')$, where $D'$ is the neighbor of $D$ in IoT-DBN. But giving a unique identity for each $ds$ may be challenging. If not, then, we have $((a: v), X)$ as the routing table entry (which is what we adopted in Section 3). Since multiple data streams may have common attribute-value pairs (descriptors), the latter has the effect of compressing the routing table. Thus, we call the latter CAC (common attribute based compression) and the former nCAC (no CAC).

In contrast to nCAC, CAC may result in "misleading routing". For example, consider the example given in Fig. 2. There are 3 data streams, $ds_1$, $ds_2$, and $ds_t$, hosted by three nodes, $D_1$, $D_2$, and $D_t$, in the network, where $AN^{ds_1} = ((a_1: v^w), (a_2: v^x))$, $AN^{ds_2} = ((a_1: v^y), (a_2: v^z))$, and $AN^{ds_u} = ((a_1: v^w), (a_2: v^z))$. ($D_3$ and $ds_3$ are ignored here.) $D_1$ and $D_2$ send the descriptors of $ds_1$ and $ds_2$ to node $D_4$. $RT^{D_4}$, thus, will have $(a_1: v^w)$, $(a_2: v^z)$, and other descriptors, which are further propagated to some intermediate nodes and reach $D_r$. Assume that $D_r$ receives a query $q$, where $A^q = ((a_1: v^w), (a_2: v^z))$. This query will be forwarded to both $D_4$ and $D_u$. Forwarding $q$ toward $D_4$ wastes bandwidth (misleading routing information due to CAC compression).

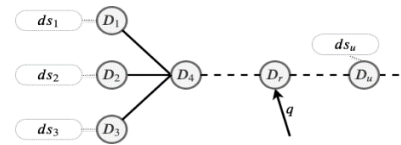

Fig. 2. *Sample of network ontology*

We conduct experiments to evaluate the impact of misleading routing in CAC. The results (will be presented in Section 8) show a significant benefit in routing table size reduction without too much messaging overhead. Therefore, we adopt CAC and will only use $((a: v), X)$ in routing table.

### 4.2 Hierarchical Based Compression

The second direction for compressing the routing table is to summarize the descriptors in a similar way as the wild card in IP address summarization. IP addresses can be summarized into common prefixes.

In general, attribute values of descriptors may be keywords or numerical values. Summarizing numerical values can be by union of numerical ranges like in [5]. We focus on summarization for attributed keywords. Also, for the MAA model, matchmaking for values of different attributes should be considered differently, so we should only summarize keywords of the same attribute. We consider different summarization policies: alphabetical based policy ($SP_{alph}$), hash based policy ($SP_{hash}$) and meaning based policy ($SP_{meaning}$). We use $SP_x$ to refer to either of these policies when discussing some strategies that are common for them.

The fundamental element in routing table summarization is which descriptors can be summarized and what they can be summarized into. To facilitate the definition of these summarization relations, we define the concept of the summarization tree. Let $ST_x(AV_{a_i}^{sys})$ denote the summarization tree (sum-tree) constructed from the set of all the descriptors in the IoT-DBN, $AV_{a_i}^{sys}$, for attribute $a_i$ using summarization policy $SP_x$ (sometimes abbreviated as $ST_x$). In $ST_x$, the child nodes (st-children) are candidates to be summarized into their parent node (st-parent) and summarization can be done recursively, from leaves to root.

$SP_{alph}$ is a straightforward extension of IP summarization. In $ST_{alph}$, an st-parent "string" is the longest common prefix (LCP) of its st-children "strings".

$SP_{hash}$ is designed to probabilistically improve the compression performance of $SP_{alph}$. Compression in $SP_{alph}$ depends on the probability of prefix match among a set of keywords in a routing table, which further depends on the keyword distributions in the application domain. We use hashing to map keywords to a uniform space. In $ST_{hash}$, an st-parent "hash code" is the LCP of its st-children "hash codes".

$SP_{meaning}$ considers the keyword distributions in the topology of IoT-DBN. IoT data streams collected by the sensors of a specific system, or a specific environment may have attributed keywords that are semantically similar. If we summarize routing table entries based on the semantic meaning of the attributes, we may get better compression in some neighborhoods. For example, some cities are major manufacturing sites for certain products, such as Detroit for automotive manufacturing, Chicago region for iron and steel industry, etc. Specific sensor data streams for the production line monitoring for these specific industries may be generated. For instance, keywords like engine-speed, engine-pressure, air-fuel, crank-position, throttle-position, etc., are likely to appear in automotive manufacturing regions and they may be "summarized" due to their relevance and semantic similarities. Fig. 3 illustrates the region dependent distribution of keywords that may be better summarized by their semantic concepts.

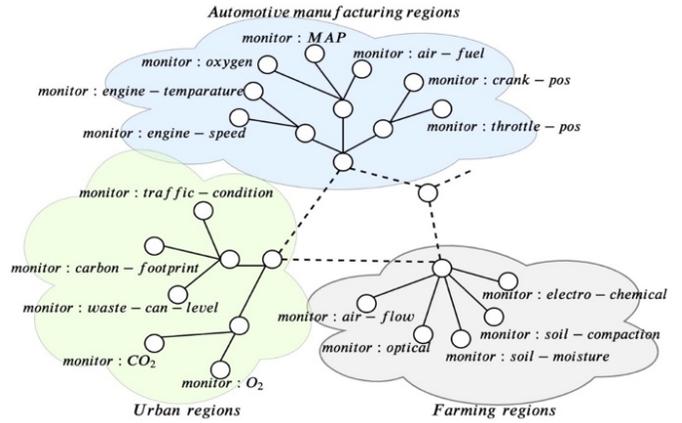

Fig. 3. *Region dependent distribution of keywords*

**Summarization trees**. The summarization tree maintains the parent-children relations of summarization, i.e., children nodes are to be summarized into the parent node. Also, instead of the raw attribute values (keywords), each node $t$ in $ST_x$ maintains a "code", denoted by $t.\tau_i$, which is derived from $t.v_i \in AV_{a_i}^{sys}$ or from the code(s) of its parent or children in $ST_x$. Let $SC_{a_i}(t^p) = \{t_j \mid 1 \leq j \leq t^p.nc\}$ denote the set of $t^p.nc$ children of a parent node $t^p$. Also, let $SCC_{a_i}(t^p) = \{\tau_{i,j} \mid 1 \leq j \leq t^p.nc\}$ denote the set of codes maintained by st-children in $SC_{a_i}(t^p)$ and $\tau_i^p$ denote the code of $t^p$, i.e., $\tau_{i,j} = t_j.\tau_i$ and $\tau_i^p = t^p.\tau_i$. As can be seen, $\tau_{i,j}$, for all $j$, can be summarized into $\tau_i^p$.

$ST_{alph}$. In $ST_{alph}$ for attribute $a_i$, a leaf node with value $v_i$, $v_i \in AV_{a_i}^{sys}$, has code $\tau_i = v_i \cdot \$$, i.e., concatenating terminator '\$' to its value. Also, $\tau_i^p = LCP(\tau_{i,j})$, $1 \leq j \leq nc$, where $LCP$ is the longest common prefix function. As can be seen, $ST_{alph}$ is a trie [17]. Note that without the terminator, we cannot differentiate a keyword from a summarization code. For example, consider a node with $\tau_i = $ CO and it has children $CO_2$ and $CO_3$. We will not know whether CO is a keyword in $AV_{a_i}^{sys}$ or just the code of an internal node in $ST_{alph}$.

$ST_{hash}$. In $ST_{hash}$, let the depth of the tree be $d$ and the maximal degree for any node be $2^c$. We have $(2^c)^d \geq \|AV_{a_i}^{sys}\|$. A leaf node $t$ of $ST_{hash}$ has code $t.\tau_i = h(t.v_i)$, where $h$ is a hash function mapping each attribute value into a code of $c \times d$ bits with a 1 added as the left-most-bit (a total of $c \times d + 1$ bits). Also, $\tau_i^p = \lfloor \frac{\tau_{i,j}}{2^c} \rfloor$, $1 \leq j \leq nc$. The code of each st-parent is constructed by removing the last $c$ bits from the code of any of its st-children.

In $ST_{hash}$, $c$ and $d$ are configurable parameters. Larger $c$ and $d$ result in a sparser summarization tree, making compression less effective. On the other hand, smaller $c$ and $d$ results in a smaller hash space, limiting the extensibility of the coding scheme. Hashing could incur collisions, but it is equivalent to preliminary summarization.

$ST_{meaning}$. $ST_{meaning}$ is constructed by two operations, clustering and coding. First, we apply Word2Vec to map keywords in $AV_{a_i}^{sys}$ to the corresponding set of vectors, denoted as $w2v(AV_{a_i}^{sys})$. $w2v(AV_{a_i}^{sys})$ is considered as the *root* vector set, and it is coded by $root.\tau_i = 1$. Then, we cluster the root set into $2^c$ clusters of vector sets using the K-mean algorithm and each cluster becomes a child of the root. We recursively cluster each parent set with code $\tau_i^p$ into $2^c$ child clusters and the code of its $j$-th st-child will be $\tau_{i,j} = \tau_i^p \times 2^c + j$. Clearly, the resulting tree $ST_{meaning}$ keeps semantically similar keywords in the same subtree.

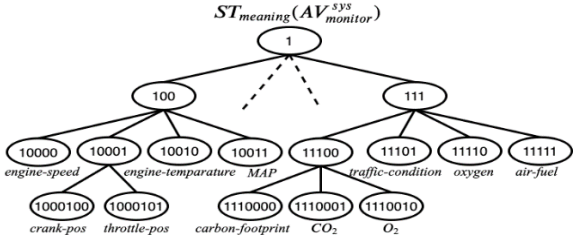

Fig. 4. Sample of $ST_{meaning}$

Fig. 4 shows the coding of an example $ST_{meaning}$ with $2^c = 4$. First, all keywords belong to the $root$ node and $root.\tau_i$ is coded by 1. These keywords are clustered into 4 child nodes and their codes are generated by appending 00 to 11 to the parent code, i.e., $\tau_i$=100 to 111. The clustering process continues till a node has only $k$ keywords in its cluster and $k \leq 2^c$. If $k = 1$, then the node becomes a leaf node. If $1 < k \leq 2^c$, then the node generates $k$ children without clustering. Thus, each leaf node has only one keyword. Each child node appends its index to its parent code to obtain its $\tau_i$ value. Thus, nodes at levels 2, 3, and 4 are coded by 3, 5, and 7 bits, respectively.

**Make summarization for keywords feasible by summarization tree based coding**. If we maintain raw attribute values $(a_i, v_i)$ in routing tables, we will not be able to identify which child keywords can be summarized into which parent keyword unless we maintain $ST_x$ at each node in the network. But this will fully defeat the purpose of summarization. It will also make per node matchmaking very inefficient. The solution is to code $ST_x$ such that the parent-children summarization relations are carried by the code $(a_i, \tau_i)$. As discussed above, the coding design in our $ST_x$ can be used to index the routing tables to (1) enable summarization, (2) further reduce routing table size, and (3) support efficient matchmaking.

For $SP_{meaning}$ and $SP_{hash}$, a group of codes of length $len$ bits are siblings in $ST_x$ if they have a common prefix of length $len - c$. For example, in Fig. 4, codes 1000100 and 1000101 have the same prefix 10001 and they can be summarized into 10001. In $SP_{alph}$, the codes are keywords and, naturally, siblings are identified by their common prefix strings.

The code size constructed via $ST_{hash}$ and $ST_{meaning}$ are very compact. With 100K keywords and $2^c = 4$, we need a tree with $d = 9$ and code $t.\tau_i$ will be 19 bits.

**How can a node in IoT-DBN know the codes of its descriptors**? When new data streams are created at a node $x$ in IoT-DBN, all descriptors $(a_i: v_i)$ are transformed into $(a_i, \tau_i)$ and used in advertisement. When issuing a query, the querier also needs to map the attributed keywords in the query to the corresponding codes. For $SP_{alph}$ and $SP_{hash}$, $x$ can derive the code $\tau_i$ directly from the attribute value $v_i$ as discussed above. $x$ does need to know $c$ and $d$ values, which can be obtained from a neighbor when $x$ joins the IoT-DBN. For $SP_{meaning}$, we have to construct the tree $ST_{meaning}$ to derive the code for each attribute value. We cannot maintain $ST_{meaning}$ on each node, thus, we use a central summarization tree server $STS$ to construct and host $ST_x$. IoT-DBN peers can query $STS$ to get the code mapping for the descriptors it hosts. We can also replicate $ST_{meaning}$ on multiple super-peers to improve availability, distribute the load, and reduce communication cost for code mapping queries. Since code mapping queries are only done upon the creation of new data streams and discovery queries, the overhead will be reasonable.

## 5 WHEN TO SUMMARIZE

Assume that an st-parent $t^p$ has an st-children set $SC_{a_i}(t^p)$ of size four. In Section 4, we summarize when all four st-children code in $SCC_{a_i}(t^p)$ are present in a routing table $RT^D$ for the same neighbor. However, if the probability for getting such "Full Sibling Code Set" (FSCS) in each routing table is not sufficiently high to achieve the desired table compression rate, then we consider more aggressive summarization. Specifically, let $cov$ denote the "coverage threshold", $0 \leq cov \leq 1$. For a given st-children code set $SCC_{a_i}(t^p)$, if $RT^D$ only has $rc$ entries that match the codes in $SCC_{a_i}(t^p)$, we still summarize them to the corresponding st-parent code $\tau_i^p$ as long as $rc \geq cov \times \|SCC_{a_i}(t^p)\|$.

Summarizing without full FSCS can also cause misleading discovery query routing. Consider $SCC_{a_i}(t^p) = \{\tau_{i,j}, 1 \leq j \leq 4\}$. Also, consider 4 data streams $ds_1, ds_2, ds_3$, and $ds_u$, where $AN^{ds_1} = ((a_i: v^w))$, $AN^{ds_2} = ((a_i: v^x))$, $AN^{ds_3} = ((a_i: v^y))$, and $AN^{ds_u} = ((a_i: v^z))$. The code for attribute values $v^w$, $v^x$, $v^y$, and $v^z$ in $ST_x(AV_{a_i}^{sys})$ are $\tau_{i,1}, \tau_{i,2}, \tau_{i,3}$, and $\tau_{i,4}$, respectively. Consider the network topology in Fig. 2. Assume that nodes $D_1, D_2, D_3$, and $D_u$ host $ds_1, ds_2, ds_3$, and $ds_u$, respectively. $D_1, D_2$ and $D_3$ send the descriptors of $ds_1, ds_2$ and $ds_3$ to $D_4$. $RT^{D_4}$, thus, will have $(a_i: \tau_{i,1})$, $(a_i: \tau_{i,2})$ and $(a_i: \tau_{i,3})$. Assume that the coverage threshold $cov$ is set to 0.7. Following the summarization rule, $D_4$ will summarize $(a_i: \tau_{i,1})$, $(a_i: \tau_{i,2})$ and $(a_i: \tau_{i,3})$ into $(a_i: \tau^p)$, which is propagated to several intermediate nodes and reaches $D_r$. When $D_r$ receives a query $q$, where $A^q = ((a_i: v^z))$, which is converted into $(a_i: \tau_{i,4})$, misleading routing will happen i.e., $q$ may be routed to $D_4$ and get no result due to the misleading summarized descriptor $(a_i: \tau^p)$. Thus, the choice of coverage threshold $cov$ has to be done carefully. We will experimentally evaluate the impact of $cov$ to come up with a good $cov$ setting.

No matter whether we consider $cov = 1$ or a lower threshold, we need to know the size of the FSCS in order to make summarization decisions. But we do not want to store $ST_x$ on each node to provide this information. Thus, we develop methods to provide FSCS size information and they are discussed in the following subsections.

### 5.1 FSCS Size Vector

To maintain the accurate FSCS size for summarization, we can associate the size information to each descriptor $(a_i: \tau_i)$. The only way to obtain the FSCS size is from the summarization tree by counting the siblings. To avoid needing to host $ST_x$ on each IoT-DBN node, we use the same server $STS$ discussed in Section 4 to provide the size information.

**Sibling count vector (SCV)**. Let $t.ns$ be the number of siblings of node $t$ in $ST_x$. $t.ns = t^p.nc - 1$, if $t^p$ is the parent of $t$. For $ST_{meaning}$ and $ST_{hash}$, since each st-parent can only have at most $2^c$ st-children, $t.ns$ can be specified by $c$ bits. For $ST_{alph}$, if the number of all possible characters in $AV_{a_i}^{sys}$ is $N_{char}$, then $t.ns$ can be specified by $\lceil \log_2 N_{char} \rceil$ bits.

An IoT-DBN peer can query $STS$ to obtain the sibling counts for its data stream descriptors. However, if summarization has taken place and the summarized parent entry is added to the routing table, we will need to contact $STS$ to get its sibling count. To avoid such communication overhead, each node $t$ in $ST_x$ maintains a sibling count vector $t.scv$, which contains not only $t$'s sibling count, but also the sibling counts of all $t$'s ancestors in $ST_x$. Based on the requirement, $t.scv$ can be constructed by composing the $ns$ values of all its ancestors. More formally, we can construct $t.scv$ by $t.scv = (t^p.scv) \cdot (t.ns)$, where $t^p$ is the st-parent of $t$. If $t$ is the root, $t.scv = \varepsilon$ (empty string). Note that for $ST_{meaning}$, we only need to maintain $t.scv = t.ns$ for the leaf nodes because each internal node always has $2^c - 1$ siblings.

SCV is very space efficient. In $ST_{hash}$ and $ST_{meaning}$, $t.scv$ is of size $\|t.\tau_i\| - 1$. In $ST_{alph}$, it is of size $\lceil \log_2 N_{char} \rceil \times len$, where $len$ is the length of the keyword. When an IoT-DBN node advertises a descriptor $(a_i : \tau_i)$ to its neighbors, the associated SCV will be carried along with the descriptor. After summarization, the SCV of the st-parent can be derived by removing the last $y$ bits of the SCV of any st-children, where $y = c$ for $ST_{hash}$ and $y = \lceil \log_2 N_{char} \rceil$ for $ST_{alph}$. For $ST_{meaning}$, the SCV of the st-parent, after summarization, will always be $2^c - 1$.

## 5.2 Average FSCS Size Estimation

When it is not desirable to communicate with a central server to obtain $t.scv$, we can develop a mechanism for estimating the number of siblings for each node $t$, i.e., estimate $t.ns$. Note that in $SP_{meaning}$, for each data stream descriptor, we anyway have to retrieve its code from $STS$, we can by the way retrieve its SCV. So, there is no reason to consider $t.ns$ estimation for $SP_{meaning}$. In the following, we consider $t.ns$ estimation for $ST_{hash}$ and $ST_{alph}$.

**For $ST_{hash}$**. In $ST_{hash}$, $t.ns$ depends on the choice of $c$ ($2^c$ is the bound of $t.ns$ for any $t$ in $ST_{hash}$), the number of leaf nodes in $ST_{hash}$ ($= \|AV_{a_i}^{sys}\|$), and the hash space size. Since the leaf nodes in $ST_{hash}$ are generated in the range of $[(2^c)^d, (2^c)^{d+1} - 1]$, the hash space size is $(2^c)^{d+1} - (2^c)^d$. Thus, the probability that a slot in the hash space has a leaf tree node is $\omega = \|AV_{a_i}^{sys}\| / ((2^c)^{d+1} - (2^c)^d)$, assuming that the hash function in use does map a keyword space to a uniformly distributed hash code space.

Since tree nodes at different levels have different expected number of siblings, $t.ns$ also depends on $t.l$, where $t.l$ is the level $t$ is at in $ST_{hash}$. Thus, we derive $f(\omega, c, l)$, which is the expected sibling set size at level $l$ and $t.ns = f(\omega, c, t.l) - 1$ ($t.ns$ is the number of siblings of $t$, excluding $t$ itself). If $f(\omega, c, l) < 1$, then $t.ns = 0$.

Let $\gamma_l$ be the existence probability of a node in a sibling set of level $l$. Then, $2^c \times \gamma_l$ will be the expected sibling set size. We have $f(\omega, c, l) = 2^c \times \gamma_l$, and $\gamma_l = 1 - (1 - \omega)^{2^{(d-l)c}}$.

To derive $\gamma_l$, we first consider the leaf sibling set, i.e., $l = d$, where $d$ is the depth of $ST_{hash}$. Obviously, $\gamma_d = \omega$. For a parent node $t_p$ of some leaf tree nodes in $ST_{hash}$, $t_p$ does not exist if it does not have any child, which has probability $(1 - \omega)^{2^c}$. Otherwise, $t_p$ does exist. Thus, $\gamma_{d-1} = 1 - (1 - \omega)^{2^c}$. Further up the tree at level $d - 2$, a node does not exist if it does not have any child at level $d - 1$, which has probability $(1 - \omega)^{2^{2c}}$ and, hence, $\gamma_{d-2} = 1 - (1 - \omega)^{2^{2c}}$. In general, we have $\gamma_{d-l} = 1 - (1 - \omega)^{2^{lc}}$.

We round $f(\omega, c, l)$ to the nearest integer to give the estimate of the number of siblings. To evaluate the accuracy of this estimation, we get the entire English keyword ontology from the dictionary of Wordnet (including 155K keywords). From the full ontology, we randomly generated 400 sub-ontologies of different sizes. The keywords from each sub-ontology are used to construct the corresponding $ST_{hash}$ using $c = 2$. We validate the estimations against the true FSCS sizes for every node in $ST_{hash}$ and obtain the mean absolute error $MAE = 0.27$ and the mean absolute percentage error $MAPE = 9.65\%$. This gives enough accuracy for the summarization decision purpose.

**For $ST_{alph}$**. Since $ST_{alph}$ does not have parameter $c$, we express the sibling set size as $f(\omega, l)$, where $\omega = \|AV_{a_i}^{sys}\|$, and $t.ns = f(\omega, t.l) - 1$. In $ST_{alph}$, FSCS size depends on the distribution of the keywords, so it is not realistic to theoretically estimate $f(\omega, t.l)$. Thus, we use the sub-ontologies generated based on Wordnet (same sub-ontologies as discussed previously) to learn $f(\omega, l)$. We collect data points $<\omega, t.l, t.ns>$ for all $t$ in $ST_{alph}$ from 80% sub-ontologies and apply various regression learning models on them, including linear, quadratic, cubic, logarithmic, rational, and exponential. The results show that $f(\omega, l)$ has almost no dependency on $\omega$. So, we derive $f(l)$ and the model with the lowest error rate is $f(l) = 26 \times l^{-2}$.

Similar to $ST_{hash}$, $f(l)$ is rounded to the nearest integer to get the estimation. We apply the estimation model to the 20% remaining sub-ontologies to validate the estimation accuracy. Error analysis shows $MAE = 0.22$ and $MAPE = 5.79\%$.

**Remarks**. To enable FSCS size estimation under $SP_{hash}$, each node in IoT-DBN needs to know the values of $c$, $d$, $\omega$ and $l$. We assume that the system is relatively stable. Thus, $c$ and $d$ can be predetermined for the network. $\omega$ may change dynamically and we assume that a growth function for the number of data streams ($\|AV_{a_i}^{sys}\|$) is given. If the system needs to update $d$ due to data growth, a leader may initiate a flooding message to inform each IoT-DBN node to change $d$ and the corresponding hash code length. The descriptor for a real data stream always has $l = d$. For FSCS size estimation for $SP_{alph}$, we only need to know $l$, where $l = length(\tau_i)$. For $SP_{hash}$, $l$ can be derived from code $\tau_i$.

# 6 THE GROWING IOT-DBN

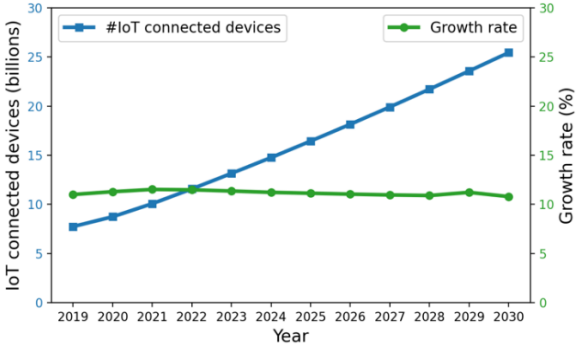

Fig. 5. *Growth of the number of IoT devices worldwide (based on prediction)*

In IoT-DBN networks, new data points are generated continuously for each data stream, but the descriptors for each existing data stream do not change. However, new IoT devices may be introduced continuously which will generate new data streams. Based on the study in Statista [19], the number of connected IoT devices worldwide has been and will be increasing by 11% per year in average from 2019-2030 (as shown In Fig. 5), i.e., the IoT network size doubles every 5 years. In fact, data analysis workflows will also generate new data streams. We can assume that these indirectly created data streams have the same growth rate as that from new IoT devices directly.

The growth of IoT-DBN has different impacts on $SP_{alph}$, $SP_{hash}$, and $SP_{meaning}$. When a new data stream is created, the new codes for its descriptors should be generated. Consider creating and handling a new descriptor $(a_i: v_i)$ that has the code $\tau_i$ in $SP_x$.

**Handling in $SP_{alph}$ and $SP_{hash}$.** In $SP_{alph}$, since the code $\tau_i$ is the prefix of $v_i$, which can be effortlessly generated by the source node. Similarly, for $SP_{hash}$, new code $\tau_i$ is the direct hash of $v_i$ and, hence, can also be generated by the source node as long as each node maintains the hash function. In both cases, routing for the new code by the nodes in the IoT-DBN network will not be an issue either.

*Increase extensively of new descriptor number.* When the number of new descriptors increases extensively, there will be a lot of collisions (same prefix in $ST_{alph}$ or same hash value in $ST_{hash}$), which will result in the growth in the number of misleading routings due to summarization coding. To cope with the growing pain and ensure a good discovery routing performance, we need to increase the code length of descriptors. For $SP_{hash}$, the code length of $\tau_i$ is related to the configurable parameters $c$ and $d$. Thus, increasing code length requires the increase of the depth $d$ of $ST_{hash}$. Meanwhile in $SP_{alph}$, since the code $\tau_i$ is the prefix of $v_i$ on each summarization tree (its length depends on the number of characters summarized on each sum-tree), we cannot directly change the code to retrieve the longer string. Therefore, changing the code requires re-summarization. In both circumstances, each node cannot autonomously change the routing table codes, and the network needs to be reestablished. In fact, when to reestablish the IoT-DBN can also be an issue. Each node will not have sufficient knowledge about the collision rate or misleading routing rate in the overall IoT-DBN.

*Distributed solution for reestablishment.* We need a distributed solution for detecting when to reestablish the network and how to do it for both $SP_{alph}$ and $SP_{hash}$. This can be done by observing the number of new descriptors. When a new data stream joins the network and starts to advertise its descriptors, super-peers will also receive the descriptors. Each super-peer maintains a copy of descriptor sets, so that it can count the number of new descriptors added to the network. When the number of new descriptors is over the threshold $T_d$ at a particular super-peer, a reestablishment of the IoT-DBN will be initialized. Reestablishment request will be forwarded to the super-peer leader and the leader will trigger the reestablishment process and send the request to all super-peers that then reset the new descriptor count and forward the request to nodes in their clusters. The second approach measures the ratio of the number of unique keywords over the hash size (for $ST_{hash}$) or over the number of leave nodes on the trie tree (for $ST_{alph}$). When the ratio is over a certain threshold, the reestablishment process for IoT-DBN is triggered. A third approach can simply reestablish the IoT-DBN network periodically according to some pre-estimated period.

Upon reestablishment, all data stream owners shall re-advertise their descriptors and each routing node shall update their routing tables. In $SP_{hash}$, hash functions will be updated with larger $d$ (also the depth of the sum-tree). The new $d$ will be sent from super-peers to each node along with the re-advertise request to generate new codes $\tau_i$ for keywords. For $SP_{alph}$, recreating routing tables will infer the larger sum-trees since it reduces the number of summarizations. The codes $\tau_i$ for this policy remain the same. The establishment process can help to decrease the misleading rate and eliminate the unused data streams if they do not re-advertise their descriptors.

*Reduce the frequent reestablishment.* Reestablishment of IoT-DBN may requires a significant effort in the network. To avoid the need for frequent reestablishment in IoT-DBN, in $SP_{hash}$, we keep a larger code size to allow better tolerance of the growth. We can use a larger hash space, i.e., generate a longer hash code size. However, a hash space that is significantly larger than $\|AV_{a_i}^{sys}\|$ may result in bad routing table compression. In our solution, we generate a longer hash code to support the growth in a period of time, then we reestablish the network when necessary. For example, if we have 100K descriptors, we may want to only use 19 bits for hash code $\tau$. But we can instead generate hash codes of 21 bits and the system can offer 4 times of the hash space without having to change the existing codes in the routing tables (23 bits implies tolerance of 16-fold of growth). Since the growth rate of the number of IoT devices is around 11%, a 4-time larger hash size can work without reestablishing the network in a couple of years.

**Handling in $SP_{meaning}$.** For a new data stream $ds$, we need to generate new codes for all its descriptors and this step is more complex for $SP_{meaning}$. Consider a new descriptor $(a_i: v_i)$. Similar to the construction of $ST_{meaning}$, we first map $(a_i: v_i)$ to the Word2Vec representation $(a_i: w2v(v_i))$. Then, we hierarchically cluster $(a_i: w2v(v_i))$.

Starting from the top layer of $SP_{meaning}$, we compute the distance between $(a_i: w2v(v_i))$ with each centroid of the child clusters and decide which cluster to go further based on the distance. This is done recursively till we reach the cluster $CL$ at the bottom layer, where all children of $CL$ are leaf nodes. If cluster $CL$ is full (i.e., has $2^c$ children) and re-clustering is not desirable yet, then we perform similarity based matchmaking to find the node $(a_i, w2v(v_j))$ that is most similar to $(a_i, w2v(v_i))$ in cluster $CL$, and the code $\tau_j$ of $(a_i, v_j)$ will be used for $(a_i: v_i)$, i.e., $\tau_i = \tau_j$. Otherwise, $\tau_i$ will be assigned to the smallest available code in $CL$.

As can be seen, the procedure discussed above requires the knowledge of $ST_{meaning}$, which is not hosted by the routing nodes. We consider the hierarchical super-peer network structure as discussed for $SP_{alph}$ and $SP_{hash}$ and use it for $SP_{meaning}$. Each super-peer hosts the entire $ST_{meaning}$ and the whole set of data stream descriptors $AV_{a_i}^{sys}$. When a new data stream $ds$ is generated and processed by a source node $D_s$ in IoT-DBN, its descriptors $(a_i, v_i), 1 \le i \le n$ are sent from $D_s$ to the super-peer of $D_s$ to let the super-peer generate code $\tau_i$ for $(a_i: v_i), 1 \le i \le n$. Subsequent advertisement by the $AdvP$ protocol can be initiated by $D_s$'s super-peer.

When the number of new descriptors increases extensively, there could be a lot of collisions, which will result in the issue of misleading routing (similar to the case of $SP_{hash}$). We use the same reestablishment process as discussed earlier for $SP_{hash}$ to address this issue. Specifically, each super-peer keeps track of the number of new descriptors added to the network. When the new descriptor number is over the threshold $T_d$ at a particular super-peer, a reestablishment of the IoT-DBN can be initialized. A second method can be based on the new clustering result, that measures the ratio of the number of unique keywords over the cluster size in each cluster at the bottom layer of the tree. The super-peer will trigger the reestablishment process when the ratio is over a pre-determined threshold $T_c$. A third method is to consider how good a cluster is via the intra-cluster distance, i.e., the average distance between cluster elements and the cluster centroid. For a cluster at the top or intermediate layer (not the bottom layer), the centroids of its children clusters can be considered as its elements and the intra-cluster distance can be computed accordingly. The establishment process will be triggered if the centroid distance is over a threshold $T_{ic}$. The most straightforward approach, as the fourth method, is to periodically initiate the reestablishment process. In all cases, the root super-peer or one of the super-peers can be chosen to perform re-clustering of all the descriptors $AV_{a_i}^{sys}$ to form a new $ST_{meaning}$ tree and the new codes for all the descriptors $(a_i: v_i), 1 \le i \le n$, can be derived based on the new $ST_{meaning}$ and readvertised In IoT-DBN.

To prevent frequent reestablishment, a larger cluster size $2^c$ can be used for better growth tolerance, same as the case in $SP_{hash}$.

## 7 DISCOVERY ROUTING

### 7.1 Routing Table Design

To ensure the performance of discovery routing, we need to ensure that the routing path is close to optimal and the forwarding decision process at each node is very efficient. This means that the routing table design must facilitate efficient lookup under the space constraint. Using our code design discussed in Section 4 to index the routing table can significantly reduce the keywords comparison time. Thus, we define an entry of routing table $RT^D$ of node $D$ as $((a_i: \tau_i), NB)$, where $NB$ is the set of neighbors of $D$ in IoT-DBN.

When considering efficient lookup, the hash table is a good data structure, but its space overhead may be a concern and it is inefficient in handling summarization. A binary search tree may have too many levels to traverse. $N$-ary tree ($N > 2$) has the matchmaking inefficiency at each internal tree node. Thus, we choose to structure the routing table as a trie, which is also the best data structure for efficient handling of summarization. Though the additional internal nodes in a trie incur space overhead (for $2^c = 4$, there will be $1/3$ more nodes than those in a tree), we design the trie specially to make it much less space consuming than a regular tree.

For $SP_{hash}$ and $SP_{meaning}$, we further use a full index table at the top layer to reduce the traversal depth. We call this data structure the *hybrid-TableTrie* (hTT). For $SP_{alph}$, the number of children for each node in $ST_{alph}$ is very high, implying a low density trie, not suitable for full table indexing. Thus, a regular trie is used for its routing table.

Since lookup is based on attributed keywords, we use multiple hTTs (or multiple tries in $SP_{alph}$), one for each attribute $a_i$, to further improve lookup efficiency. (This is not needed for regular keyword based lookup.) For simplicity, we discuss the hTT design for a single attribute $a_i$, which includes a master table (MT) and a trie layer consisting of RT-tries.

**Master table**. The master table is a full index table, indexing the first $b$ bits of code $\tau_i$. Thus, it contains $2^b$ elements. Each element in the table is a pointer pointing to the root of an RT-trie in the trie layer. We decompose $\tau_i$ into the master code $\tau_i^b$ and the trie code $\tau_i^{b+}$. During decomposition, the leading 1 in the code is shifted to $\tau_i^{b+}$. Specifically, let $\tau_i = 1 \cdot \tau_i'$ and $\tau_i^{b+} = 1 \cdot \tau_i''$, then we have $\tau_i' = \tau_i^b \cdot \tau_i''$ and $\tau_i^b$ is the $b$-bit prefix of $\tau_i$ without the leading 1. The choice of $b$ depends on the size of $\|AV_{a_i}^{sys}\|$ and larger $b$ can be chosen for larger $\|AV_{a_i}^{sys}\|$. We also need to ensure that there exists no leaf node $t$ in $ST_x$ such that $t.l \le b/c$. This is especially critical in $ST_{meaning}$ since its leaves can appear at any level (in $ST_{hash}$, a leaf node is always at the lowest level). Also, summarization should not go beyond level $b$.

Same as $\tau_i$, $scv$, if maintained, can also be decomposed as $scv = scv^b \cdot scv^{b+}$. Since summarization will not go beyond level $b$, $scv^b$ can be discarded, we still refer to $scv^{b+}$ as $scv$.

Consider a routing table entry $r$. Let $r.l$ denote the level of $r.\tau$ in $ST_x$. A type A entry $r$ is likely to have its $r.l > b$. Thus, each master pointer $ptr$ points to a routing entry $r$ with $r.l = b + 1$. If an entry $r$ of type A has $r.\tau.l < b$ and $r$'s ancestors are of type P, then $ptr$ should point to $r$ and $r$ only needs to point to one child whose code is a prefix of $ptr$. However, this situation is unlikely to happened and we will not consider it in the algorithms given next.

**RT-trie**. Each RT-trie is associated to a pointer in MT. Nodes in RT-trie are routing table entries that are constructed from $\tau_i^{b+}$. To ensure no unnecessarily wasted space, we consider three types of entries, the type A *actual* entry, the type P *pointer* entry, and the type M *mixed* entry. A pointer entry is for maintaining the structure of the trie, without routing data. It simply maintains pointers, *children*, which includes $2^c$ pointers to its child entries. The actual entries are leaf nodes in the RT-trie and actually have routing information $(\tau_i^{b+}, NB, scv^{b+})$. For entry $r$, $r.NB$ is the set of neighbors for query forwarding. $r.\tau_i^{b+}$ is the code for the entry. $r.scv^{b+}$ is the FSCS size vector for $r$. A type M entry maintains $(\tau_i^{b+}, NB, scv^{b+}, children)$, including the information of both type A and type P entries. During routing table updates, the types of entries may change. A sample hTT is shown in Fig. 6.

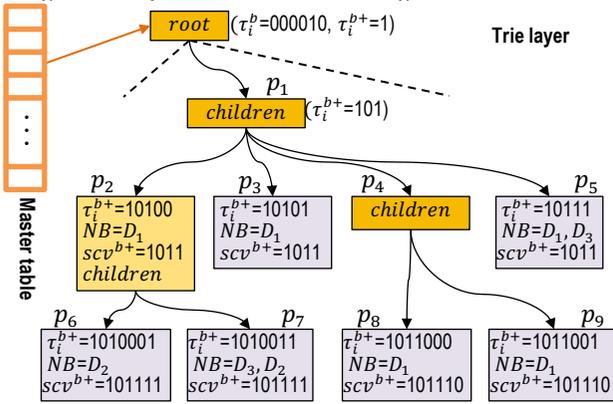

Fig. 6. *Sample of a hTT*

In the example, we consider using the $SP_{hash}$ with 13-bit code and $b = 6$. Thus, each MT entry points to a 4-level RT-trie (level 0 to 3, with *root* being level 0). The third entry in the MT has $\tau_i^b = 000010$, which points to the example RT-trie, including 6 type A entries (in the lightest color), one type M entry $p_2$, and 3 type P entries $p_1, p_4$ and *root*. $\tau_i^{b+}$ for *root* of the RT-trie is 1. $p_3$ and $p_5$ are summarized entries. From $scv^{b+}$ of $p_6$ to $p_9$, we can see that $p_2$ and $p_4$ may have upto 4 and 3 children, respectively. $p_2$ is also a summarized entry. Originally, $p_2$ had 4 children with $D_1$ in their $NB$ field and they are summarized into $p_2$. But since $p_6$ had $NB = \{D_1, D_2\}$ and $p_7$ had $NB = \{D_1, D_2, D_3\}$ and only $D_1$ is common in all $p_2$'s 4 children, so $p_6$ and $p_7$ cannot be removed.

**Detailed routing table design for large scale systems**. We consider a large-scale system with over 10K keywords in its routing table for one attribute. We choose $b = 8$ and $2^c = 4$, resulting in a master table of 256 entries, which can save 3 traversals. Each pointer in our design is of 16 bits (will be discussed later). Thus, the index table only has a small footprint of size 512B.

By having different node types, we strip off unnecessary data in each node to minimize its space. But this design requires each node to carry a 2-bit type information. We design a specific structure for the internal of each node to integrate the 2 type bits and other data in the node and to minimize the per node space requirement.

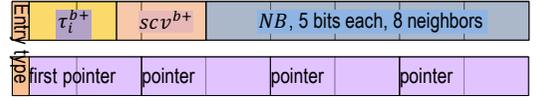

Fig. 7. *Sample design of routing table entries.*

We allocate 8 bytes to each type A or type P entry as shown in Fig. 7. The first 2 bits are needed to indicate the entry type. For a type A entry, we use 11 bits for code $\tau_i^{b+}$, 11 bits for $scv^{b+}$, 40 bits for $NB$. In this allocation, fields $\tau_i^{b+}$ and $scv^{b+}$ can support over 100K keywords (with only 10K keywords in the routing table) for attribute $a_i$ and field $NB$ can address 8 neighbors with 5-bit each. The detailed information about the neighbors, such as their IP address, should be maintained globally on node $D$. The 5-bit $NB$ code is sufficient to identify each neighbor of $D$ for up to 32 neighbors. If more than 8 neighbors send code $\tau_i^{b+}$ to $D$, the entries that arrive late will be discarded. Also, $\tau_i^{b+}$ can be omitted since it can be implied from the trie structure, and the space can be used for $NB$.

For the pointers in type P entries, instead of using the memory address pointers, which is generally of 32 bits, we maintain a heap for the routing table (RTheap) and use 16-bit pointers, each point to an 8-byte table entry. We unify the sizes for different types of table entries to 8 bytes (or double of 8 bytes for type M). The true memory address can be computed by $ptr * 8 + base$, where $base$ is the starting address of RTheap and it is only stored once in the system.

Each type P entry stores $2^c = 4$ pointers (we can double it if $2^c = 8$), each is of 16 bits. The $i$-th pointer points to the $i$-th child entry so that there are no comparisons needed during traversal. To accommodate the 2-bit entry-type field, only 14 bits will be given to the first pointer. We use a special allocation strategy, top-bottom allocation (TBalloc), to guarantee that the first pointer will only require 14 bits as long as the number of entries the first pointer may point to will be less than $1/2$ of the entries in the routing table. This is a reasonable assumption because the probability for each pointer being null is the same.

**TBalloc**. Each entry in the routing table has a code. An entry is called the zero entry (ZE) if its code ends by 00 ($\tau_i^{b+} = \cdots 00$). As can be seen, the first pointer in a type P entry will either be null or points to an ZE. We allocate ZE from the beginning of the heap first. Allocation for non-ZE entries starts from the middle of the RTheap, grows evenly along the up and down directions. If the first $1/4$ of the heap are allocated to ZEs or allocation of non-ZE entries grow beyond the $1/4$ heap boundary, then allocation of ZE continues from the bottom of the heap. The allocation layout is shown in Fig. 8. As a result, 14 bits are sufficient to address the top $1/4$ or the bottom $1/4$ of the heap space. To differentiate the top and bottom addresses, we add a P' type entry. If an entry is of type P', then its first pointer address should be converted to one that starts from the bottom of the heap.

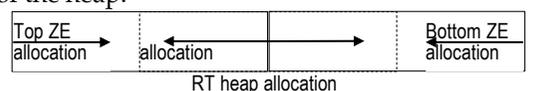

Fig. 8. *Top-bottom allocation for RTheap.*

A type M entry will have both types A and P entries, a total of 16 bytes, but the first pointer in its type P part can have full 16 bits. Also note that the design given above is a sample case. The table entries and the length of each field should be adjusted according to the specific parameters of the IoT-DBN.

**Analysis**. Consider a routing table with $\omega$ type A entries, $\omega \leq \|AV_{a_i}^{sys}\|$. If we organize them in a regular $N$-ary tree, there will be $\omega$ entries, and each entry includes routing information and pointers. While in a trie, with $\omega$ entries as the leaves, there will be $\cong \omega/(2^c - 1)$ additional entries to hold the trie structure. For $c = 4$, it is $1/3$ more entries than those in a tree. In our design, if $b = 0$, we only need $\omega$ information entries and $\omega/3$ pointers, which is $2\omega/3$ less pointers than those in an $N$-ary tree (because we do not carry node information inside a pointer entry and no pointer inside an actual entry). With $b \neq 0$, we have $2^b$ instead of $((2^c)^{b/c+1} - 1)/(2^c - 1) - 1$ pointers. For $2^c = 4$ and $b = 8$, we have 256 pointers instead of 340 pointers, i.e., 84 less pointers (if the trie is full at the top 4 levels). Also, the space for our pointers is minimized (2 bytes). Thus, overall, our trie design has very high space efficiency for routing table storage, even considering large-scale IoT-DBN.

## 7.2 Routing Table Algorithms

We develop algorithms for routing table accesses, including the query processing, advertisement, and summarization algorithms.

**For QFP**. When processing a query $q$, descriptors with different attributes will be processed separately. For descriptor $(a_i : \tau_i)$ in $A^q$, we look for all neighbors with the corresponding code $\tau_i^{b+}$ in the RT-trie that is pointed to by the $\tau_i^b$ entry in MT. The look up function traverses top-down from the $root$ of the RT-trie via the path on which each entry matches the prefix of $\tau_i^{b+}$, until reaching the entry with the longest prefix match. When traversing an entry $r$, if $r$ is of type A (or M), all neighbors in $r.NB$ are considered as forwarding neighbors for $\tau_i$ (could be a neighbor in a summarized entry) and are added to the forwarding neighbor set $fnset_i$ for attribute $a_i$. The look up algorithm is given in Algorithm 1. From output $fnset_i$, for all $i$, we can count the appearance of each candidate neighbor $D_i$ and determine whether $D_i$ is a $\alpha$-matching neighbor. Subsequently, the $\alpha$-matching neighbor set $mns$ (discussed in Section 3) for $q$ can be derived.

---

**Algorithm 1. QFP**
Input: $\tau_i$, output: $fnset_i$:
  From MT, get $r \leftarrow$ root of the RT trie for $\tau_i^b$;
  While $r$ is not $null$ do
    If $r$ is of type A or M, then add nodes in $r.NB$ to $fnset_i$;
    $r \leftarrow$ the $j$-th child of current $r$ if $j =$ the next two bits of $\tau_i^{b+}$;
  Endwhile

---

For example, in Fig. 6, when we look up a descriptor with $\tau_i = 1000010010001$ for query $q$ ($\tau_i^{b+} = 1010001 = p_6.\tau_i^{b+}$), we get $fnset_i = \{D_1, D_2\}$.

**For AdvP**. When an IoT-DBN node $D$ receiving an advertisement message $advmsg$ from neighbor $D'$, it checks each descriptor $(a_i : \tau_i)$ in $advmsg$ to determine whether $(a_i : \tau_i)$ is in $RT^D$. From $\tau_i^b$, We get the pointer in MT and reaches the $root$ of the designated RT-trie. We then traverse RT-trie top-down until reaching the entry $r$, which has the longest prefix match with $\tau_i^{b+}$. If $r$ is type A (or M) and $D' \in r.NB$, no action is needed and no further forwarding. If $r.\tau_i^{b+} = \tau_i^{b+}$, then we add $D'$ into $r.NB$. Otherwise, we insert a type A leaf entry $r'$ with content $(\tau_i^{b+}, \{D'\}, scv^{b+})$ together with some potential type P pointer entries into the trie as descendants of $r$ using the standard trie insertion algorithm. During RT-trie traversal, we also derive $fnset_i$ for $\tau_i^{b+}$ as in Algorithm 1. After updating $RT^D$, descriptor $(a_i : \tau_i)$ will be forwarded to all neighbors of $D$ besides those in $fnset_i$ (descriptors to the same neighbor will be packed in one $advmsg$ for forwarding). Details of the algorithm is given in Algorithm 2.

---

**Algorithm 2. AdvP**
Input: $\tau_i$, $D'$, $scv$
  From MT, get $r \leftarrow$ root of the RT-trie for $\tau_i^b$;
  # Find the longest matching prefix entry
  While $j =$ the next two bits of $\tau_i^{b+}$ and $j$-th child of $r \neq null$ do
    $r \leftarrow$ the $j$-th child of current $r$
  Endwhile
  # Add the neighbor to the entry
  If $r$ is type A or M and $D'$ is in $r.NB$ then exit;
  If $(r.\tau_i^{b+} = \tau_i^{b+})$ then add $D'$ to $r.NB$
  Else
    # Add type P entries till the location of the actual entry
    While $level(r) < d - b/2 - 1$ do
      Add a type P entry $rc$ to $r.children$
      $r \leftarrow rc$
    Endwhile
    # Add the actual entry
    Add a type A entry $rc$ to $r.children$
    $rc \leftarrow (\tau_i^{b+}, \{D'\}, scv^{b+})$
Endfunction

---

For instance, in Fig. 6, when neighbor $D_1$ advertises a descriptor $(a_i : \tau_i)$, where $\tau_i = 1\,000010\,011010$ and $scv^{b+} = 101110$, a new node $p_{10}$ with a type A entry $(1011010, \{D_1\}, 101110)$ will be inserted to $p_4.children$.

**Summarization**. Consider the routing table $RT^D$ of an IoT-DBN node $D$. When an entry $rc$ is added to $RT^D$, summarization can be done by checking whether $rc$'s siblings can be summarized into $rc$'s parent $r$. Summarization may cause misleading routing. To avoid this problem, we can also perform summarization when the routing table exceeds a prespecified size bound. In this case, we try to compress each RT-trie pointed to by the MT of each attribute. For each RT-trie, we can use recursion to traverse the trie for summarization. We develop the function $Summarize(r)$ (given in Algorithm 3) to support either of the two approaches discussed above and $Summarize(r)$ summarizes the children of a parent RT-trie entry $r$.

To achieve the compression effect, summarization should only be done for an entry $r$ when it has at least one leaf child entry. Specifically, all children of a to-be-summarized entry $r$, denoted as $rc_k$, for all $k$, should be of type A or M and at least one of them should be of type A. Also, when all the children of an entry $r$ are summarized into $r$ and removed from the RT-trie, $r$ becomes the leaf node and may be summarized further up. In this case, $r$ could be of type P or M. After summarization, if $r$ becomes a leaf entry, then it will be converted into a type A entry.

**Algorithm 3. Summarize**
Input: $r$, $cov$
  $rc_k \leftarrow$ $k$-th child of $r$ from $r.children$
  $UNB \leftarrow$ union of $rc_k.NB$
  Foreach $D_j \in UNB$, get $D_j.count \leftarrow$ #times $D_j$ appears in $rc_k.NB$
  $nc \leftarrow$ (last two bits of $rc_k.scv$) + 1
  If $D_j.count \geq nc * cov$ then add $D_j$ to $sumNB$
  Foreach $D_j$ in $sumNB$
    For all $k$, remove $D$ from $rc_k.NB$
    For all $k$, remove $rc_k$ if $rc_k$ is of type A and $rc_k.NB$ is empty
    Add $D_j$ to $r.NB$
  If $r.children$ is empty, then convert $r$ to type A

In the algorithm, we scan through $rc_k$, for all $k$, to determine the union of $rc_k.NB$ (denoted as $UNB$) and the number of appearances of each neighbor $D_j \in UNB$ in all siblings (denoted as $D_j.count$). $sumNB$ is the set of all can-be-summarized neighbors, i.e., if $D_j \in sumNB$, then $D_j.count \geq nc * cov$, where $nc$ the number of children $r$ has. After summarization, $D_j \in sumNB$, for all $j$, are removed from $rc_k.NB$, for all $k$. If $rc_k.NB = \emptyset$ and $rc_k$ is of type A, then $rc_k$ can be removed from the trie.

Consider an example. Assume that after neighbor $D_1$ advertises a descriptor ($a_i$: 1 000010 011010 ), $p_{10}$ is added to the RT-trie, and $RT^D$ exceeds its size bound. Then, summarization is triggered with $cov = 1.0$. $D_1$ in $p_8$, $p_9$, and $p_{10}$ is summarized to parent node $p_4$, and $p_8$, $p_9$, and $p_{10}$ are removed from the tree. Since $p_4$ does not have any children, it will be switched to a type A entry. After that, $D_1$ in $p_2, p_3, p_4$, and $p_5$ is summarized to parent node $p_1$. Later, nodes $p_3, p_4$ are eliminated from the tree while $p_1$ is switched into a type M entry, $p_1.NB = \{D_1\}$. Thus, 5 nodes are eliminated from this summarization, resulting in great space saving.

## 8 EXPERIMENTAL STUDY

We simulated an IoT network using the technique of Fast Network Simulation Setup (FNSS) toolchain [18]. Networks of different sizes (#nodes), from 1K to 30K nodes, are generated. Each node in the network can have 2 to 10 neighbors (following a weighted uniform distribution). All experiments were run on a Windows server with 8 Intel Xeon 3.7GHz processors, 32GB RAM. We crawled the web and obtained about 100K IoT data stream annotations that include 15 attributes and 55K unique keywords (879K keywords in total). The raw data can be downloaded from [7].

Data descriptors for each object, with some random replications, are uniformly randomly distributed to the nodes near and at the edge of the network. A query is generated by uniformly randomly selecting a data stream and uniformly randomly selecting some of its descriptors to include. For each query, a uniformly randomly selected node in the network is designated to initiate it.

We implemented AdvP and QFP to study the effectiveness of various summarization algorithms in reducing routing table size and their impact to discovery query routing performance. We also compare the performance of representative algorithms in major p2p data discovery approaches, including (1) our solutions and (2) the GSD-like peer-to-peer algorithm. Two simple DHT-based approaches are also considered, including (3) the psKey technique [1] and (4) the Mkey algorithm [12]. The case of (2) represents a set of unstructured routing algorithms that do not summarize for keywords, including GSD [4] and CBCB [5]. For fairness, the information cache for GSD is already built up before discovery queries are issued. (3) and (4) have been discussed in Section 2. For (3), we derive all the subsets of the set of keyword-based descriptors for each data stream and hash them to the DHT. In (4), the Mkey structure, which is a hybrid of DHT and unstructured cluster, is simulated. Bloom filters of size 128 bits is used to code the descriptors of each data stream and is split into hash codes for DHT storing the data stream indices.

### 8.1 Configuring Summarization Algorithms

We study the impact of the parameter settings in various algorithms, including (a) the coverage threshold $cov$, (b) the degree bound for the summarization trees, $2^c$. The impact is measured in terms of (i) RT-size (the number of routing table entries); (ii) Traffic (the average number of messages, i.e., the total number of hops); and (iii) Latency (the hop number for each query to successfully get its responses from all the routing directions). The hop bound for advertisement $b^{ad}$ and for query forwarding $b^q$ are set to $\infty$.

**(a) coverage threshold**. Coverage plays an important role for summarization. We evaluate the impact on RT-size, Traffic, and Latency due to different coverage thresholds $cov$ in various summarization policies considering different network sizes. Other parameters are set to $2^c = 4$ and $\alpha = 1.0$. The results are shown in Fig. 9, 10, and 11.

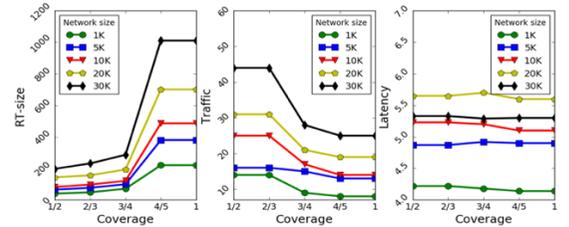

Fig. 9. *Coverage Impact for $SP_{meaning}$*

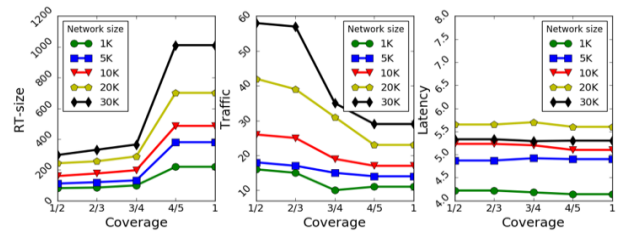

Fig. 10. *Coverage Impact for $SP_{hash}$*

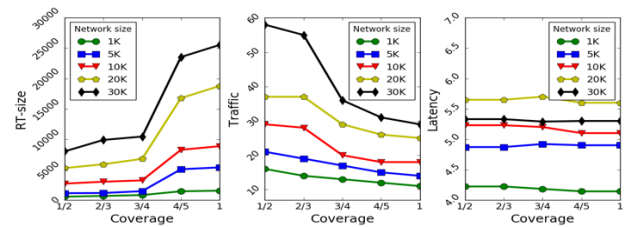

Fig. 11. *Coverage Impact for $SP_{alph}$*

Clearly, $cov = 3/4$ gives the best tradeoff between RT-size and Traffic for all policies. When $cov \geq 3/4$, RT-sizes are very reasonable, but Traffic becomes significantly higher. When $cov < 3/4$, Traffic is good, but RT-size rises sharply. Although $cov = 3/4$ yields the best tradeoff, we use $cov=1$ in experiments to show the effectiveness of summarization.

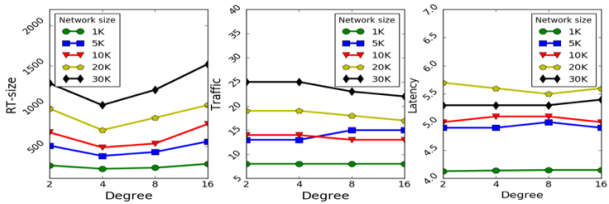

Fig. 12. *Degree comparison for $SP_{meaning}$*

**(b) Degree bound for the summarization trees**. We study the impact of different settings of $2^c$ and execute AdvP and QFP with different summarization policies and different network sizes. The other parameters are set as $cov = 1.0$ and $\alpha = 1.0$. Same reason of space limitation as above, on the results for $SP_{meaning}$ is presented in Fig. 12.

As can be seen, when $2^c = 4$, routing table has the most effective compression. RT-size rises when $2^c > 4$ because the probability of having enough siblings for summarization becomes lower. For $2^c = 2$, we further investigated the reason for the higher RT-size and found that summarization happens more frequently at the leaf nodes, but less frequently at the higher layers, resulting in a lower overall compression rate compared to the case of $2^c = 4$. Traffic for increasing $c$ drops slowly with increasing $c$ and Latency only differs marginally with different $c$ settings. Thus, we choose $2^c = 4$ for all summarization policies.

### 8.2 Misleading Routing

In addition to studying the overall Traffic and Latency, we also investigate the misleading routing problem due to (a) the use of CAC indexing, (b) the coverage threshold in summarization and (c) the mechanisms for computing coverage and (d) the collision problem in $SP_{hash}$. Understanding misled traffic can help gain fine-grained insights on the impacts of the design decisions in discovery routing methods. The parameters that are fixed in this section include $2^c = 4, \alpha = 1.0$.

**(a) CAC indexing**. We study the impact of CAC compression in terms of RT-size and Misleading (the percentage of messages forwarded due to misleading routing information) caused by CAC compression. nCAC is used as the baseline for comparison. Fig. 13 shows the experimental results.

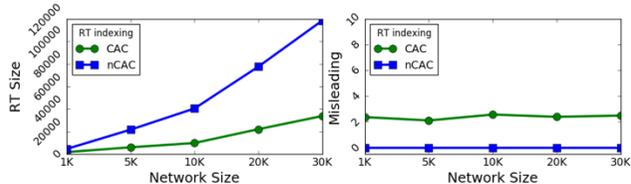

Fig. 13. *Impact of compression for CAC and nCAC*

As can be seen, using CAC will reduce the routing table size significantly, specifically, 3 to 4 times of reduction compared to nCAC. The effect is more significant with a larger network size. On the other hands, the wasted traffic due misled routing in CAC is stable and at around 2% (nCAC has no misled routing traffic).

**(b) Coverage threshold**. The percentages of misled messages in overall Traffic due to $cov \neq 1$ are shown in Fig. 14. When $cov = 1.0$, there is no misled messages due to coverage. The 2% misled messages are due to indexing by individual descriptors without identifying their data streams. For example, consider the descriptors of 3 data streams: $AN^{ds_1} = ((a_1: v^w), (a_2: v^x))$, $AN^{ds_2} = ((a_1: v^y), (a_2: v^z))$, and $AN^{ds_3} = ((a_1: v^w), (a_2: v^z))$. A node $D$ advertising descriptors in $AN^{ds_1}$ and $AN^{ds_2}$ may cause misled routing toward $D$ for queries looking for $AN^{ds_3}$. When decreasing coverage threshold $cov$, percentage of misleading message increases. $cov = 3/4$ yeilds the best tradeoff, conforming to the Traffic results in Fig. 9, 10 and 11.

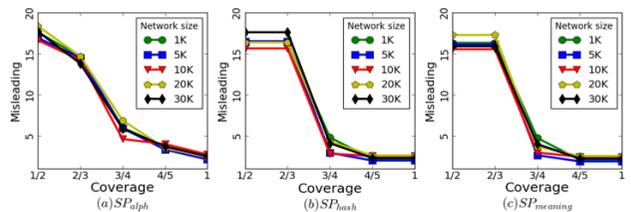

Fig. 14. *Misleading Impact for policies*

**(c) SCV vector vs FSCS size estimation**. Maintaining SCV in the routing table can incur communication costs for SCV retrieval, but it allows an accurate coverage computation. We analyzed the accuracy of FSCS size estimation and study the impact of this inaccuracy toward the percentage of misled messages and the results are shown in Fig. 15.

Since we set $cov = 1.0$, using SCV will not cause misled routing and the 2% misled messages comes from indexing by individual descriptors without identifying their data streams. Let $E$ and $A$ denote the estimated and the actual FSCS sizes. $E > A$ will not yield misleading messages. If $E < A$, coverage may be underestimated, and may result in some misleading routing. We analyzed the results and observed that 0.3% additional misleading messages are due to $E < A$, which subsequently causes the compression ratio to rise from 12.9 times to 13.5 times.

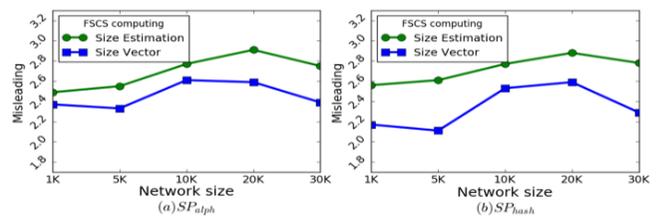

Fig. 15. *Misleading impact for FSCS computing*

**(d) Collision impacts**. In $ST_{hash}$, the number of slots available at level $d$ is $f(c, d) = (2^c)^{d+1} - (2^c)^d$, which is the hash space for hashing keywords in $AV_{a_i}^{sys}$. We generate different hash space density cases, i.e., $\|AV_{a_i}^{sys}\|/f(c, d)$, by randomly selecting a subset of keywords from the IoT dataset (for a specific attribute) as the $AV_{a_i}^{sys}$ set and build the corresponding $ST_{hash}$ tree of different depth $d$ and $2^c = 4$. We consider $\|AV_{a_i}^{sys}\|/f(c, d)$ value from $1/16$ to $1.0$. Fig. 16 shows the experimental results.

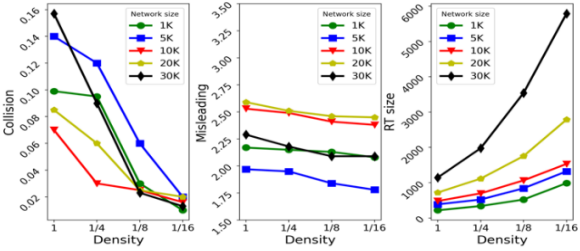

Fig. 16. *Impacts for collision in $ST_{hash}$*

The left-hand side plot shows the Collision rate (the percentage of keywords that incurs collision), and the one on the right shows the percentage of misled messages. As can be seen, $\|AV_{a_i}^{sys}\|/f(c,d)$= 1/4 to 1 yields a better tradeoff between Misleading routing rate, RT size and Collision. When the density changes from 1.0 to 1/16, number of collisions decreases significantly but the percentage of misleading messages only reduces slightly while the RT size increases noticeably at 1/8 or 1/16 level only. Thus, collision does not cause problem and a relatively dense hash space can be used to avoid the space overhead.

### 8.3 Summarization Effectiveness

We now study the effectiveness of our three summarization algorithms. We use the same random object distribution (Random) discussed in the beginning of this section as well as the regionalized object distribution (Region) discussed in Section 4 to evaluate the algorithms. For Region distribution, we automatically divide the network into several "regions" based on the traversal order. We also cluster the data streams based on the similarity of their descriptors. Then, we distribute objects in the same cluster to the same region to create the regionalized keyword distribution effect. In the experiment, we use 7 regions and 14 clusters. We use the no summarization case (called nSum) as the baseline for evaluation. We set cov=1.0. Fig. 17 shows the results.

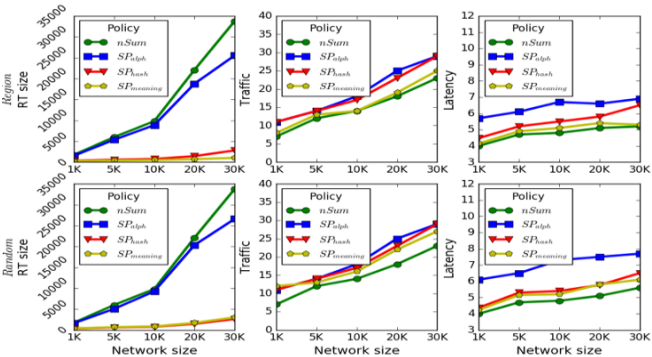

Fig. 17. *Policy comparison*

In a relatively large-scale network (30K node), $ST_{meaning}$ and $ST_{hash}$ achieves 11.1 and 12.9 compression ratios, respectively, in Random distribution, and 33.3 and 12.1 compression ratios, respectively, for Region distributions. The total Traffic increases from *nSum* due to summarization by $SP_{meaning}$ are 7% in Region and 10% in Random distribution, by $SP_{hash}$ is 12% (only considered Random), and by $SP_{alph}$ is 15%. The increase in Traffic is acceptable considering the at least 11 times reduction in RT-size. In terms of Latency, $SP_{hash}$ and $SP_{meaning}$ have roughly a 10% increase from *nSum* while the increase in Latency by $SP_{alph}$ is 30%.

Note, the comparison is based on unlimited routing table size. If we confine the RT size, some routing information may be forced to be thrown away in existing approaches, which will significantly increase their Traffic and Latency results.

Our further investigation shows that in the very large IoT-DBN of 30K nodes for Region distribution, the average RT-size per node using $SP_{meaning}$ reduces to 45KB, a significant reduction from 1249KB of *nSum*, while this number for $SP_{hash}$ and $SP_{alph}$ are 50KB and 897KB, respectively. Without using CAC indexing and any summarization policy, the average RT-size is 4121KB. In Random distribution, $SP_{hash}$ has the least routing table space, only 51KB. The detail data are provided in Table 1.

TABLE 1.
AVERAGE RT-SIZE(KB) IN 30K NODES IOT-DBN

| Techniques | nCAC | CAC(=nSum) | $SP_{alph}$ | $SP_{hash}$ | $SP_{meaning}$ |
|---|---|---|---|---|---|
| Region | 3667 | 1112 | 798 | 45 | 40 |
| Random | 3667 | 1083 | 821 | 46 | 52 |

### 8.4 Compare Data Discovery Approaches

We further compared the performance of five data discovery approaches discussed in the beginning of this section. We choose $SP_{hash}$ to represent our summarization techniques. For the GSD-like approach, we consider unbounded cache size (GSD-like-nBound) and bounded cache size (GSD-like-Bounded). In the latter case, GSD cache is bounded by the same size as the RT-size in $SP_{hash}$. The bound is achieved by using LRU to eject the extra entries. The results for Traffic and Latency are shown in Fig. 18.

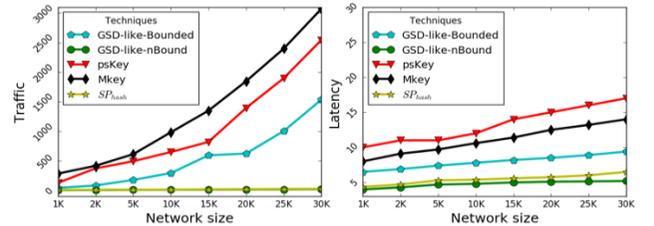

Fig. 18. *Comparison of major data discovery approaches*

As can be seen, our $SP_{hash}$ algorithm outperforms all other data discovery approaches except for GSD-like-nBound. DHT based approaches (psKey and Mkey) require routing to find the node hosting the location information and they also have wasted communications with nodes that may not have the queried results and Chord routing has inherited extra hops for each intermediate destination. Thus, they impose a much higher overhead in Traffic and Latency. In GSD-like-Bounded, useful routing information are being thrown away instead of summarized and, hence, it has worse Traffic and Latency than $SP_{hash}$. Though GSD-like-nBound solution has a slightly better Traffic and Latency than $SP_{hash}$, its cache size will be 23.5 times larger than that of $SP_{hash}$.

With increasing network size, all other algorithms besides the p2p ones, especially the DHT based approaches, have drastically increasing Traffic and Latency. At a very large scale IoT-DBN, (for instance, 30K nodes) both the p2p approaches have 2 to 4 times lower Latency and 50 to 100 times lower Traffic in comparison with psKey and Mkey approaches. GSD-like-Bounded having the same RT-size with our $SP_{hash}$ yields 20 times higher Traffic and 2 times higher Latency.

## 9 EXPERIMENTS FOR GROWING IOT-DBN

### 9.1 Network Growth Configuration

The summarization tree settings may impact the IoT network growth in different aspects. To find the best summarization tree configuration to tolerate network growth better, we conduct the experience on each policy $SP_{hash}$ and $SP_{meaning}$ with different tree configs. Since $SP_{alph}$ does not require configuration settings, we do not consider this policy for this experiment. We choose the large network with 30K nodes to demonstrate the large-scale network. Since the average growth of connected IoT devices worldwide is around 11% per year, we consider the growth rate from 10% to 100% per year, corresponding 3K to 30K nodes added to the IoT network. We randomly generate new data streams and semantically distribute them to those new nodes at the network edges. Comparison metrics, including misleading, traffic, and RT-size, are calculated after adding the new nodes.

In $SP_{hash}$, the sum tree configuration depends on c and d. Similar to previous experiments in Section 8.2(d), we generate different hash space density cases, i.e., $\|AV_{a_i}^{sys}\|/f(c,d)$, and build the corresponding $ST_{hash}$ tree of different depth $d$ and $2^c = 4$. The results in Fig. 19 show that the density of 1/4 gives the best trade-off between comparison metrics. When the density is shifted from 1/16 to 1/4, the misleading and traffic slightly increase, while the RT-size decreases. From 1/4 to 1.0, the misleading and traffic grow dramatically due to the collision in hashing. In the meantime, the RT-size decreases because of the summarization and hashing collision.

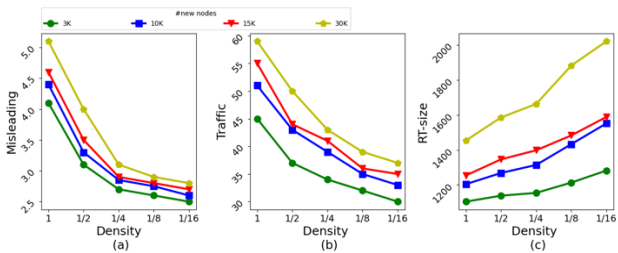

Fig. 19. $SP_{hash}$ configuration for network growth

In $SP_{meaning}$, the sum tree depends on the number of clusters $2^c$. Thus, we build the tree with different degrees of $2^c$. The results in Fig. 20 show that clustering into 4-8 clusters yields the best trade-off. The smaller degree of cluster produces a higher chance for summarization. This help to reduce the RT-size but increase significant traffic and misleading issue. Meanwhile, the larger degree leads to less summarization that impacts the RT-size but reduces the traffic and misleading rate.

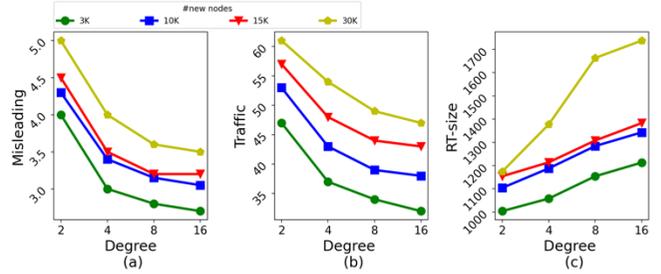

Fig. 20. $SP_{meaning}$ configuration for network growth

### 9.2 Reestablishment Trigger Analysis

In section 6, we discussed the threshold for triggering the reestablishment, and when to establish impacts the network performance. As mentioned, we measure the ratio of the number of unique keywords over the max code size capacity (for $ST_{hash}$ and $ST_{meaning}$) or the number of leave nodes on the trie tree (for $ST_{alph}$). When the ratio is over a certain threshold $T_d$, the reestablishment process for IoT-DBN is triggered. Besides, a straightforward approach is to simply reestablish the IoT-DBN network periodically according to some pre-estimated period.

To analyze the threshold $T_d$, we run experiments to evaluate the impact on RT-size, traffic, and misleading on different thresholds considering the large network of 30K nodes. The threshold changes from 1/16 to 1.0. We measure the network's performance at the point when the growing network reaches the threshold. As shown in Fig. 21, the threshold in range of [1/4 -1/2] throws the best tradeoff between the comparison metrics. When $T_d$<1/4, triggering the reestablishment may be unnecessary since the RTs still have more space for new data. When $T_d$> 1/2, more collisions happen, leading to more traffic and misleading.

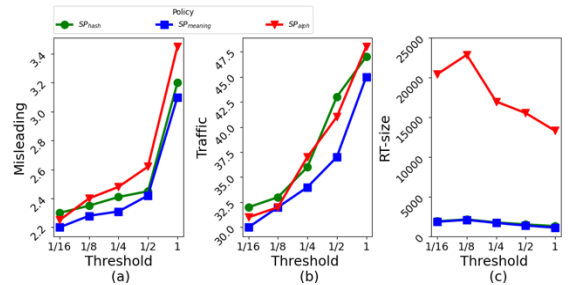

Fig. 21. *Reestablishment threshold analysis*

We also conduct the periodical reestablishment experiments with different periods. Especially, we consider the period for reestablishment from 0.5 years to 5 years, with the growing rate 11% like the actual rate of IoT networks. The metrics are computed at the end of each period. The bar charts show that the more frequent the network is reestablished, the less misleading happens, but this also increases the RT-size. The reason is that the reestablishment leads to tree capacity growth, which significantly reduces the collision rate and the number of summarization operations. Reducing the summarization by combining the network growth will enlarge the RT-size. Frequent reestablishment requires more resources. Thus, the 1-2 years period can be considered the best tradeoff between RT-size and misleading and traffic. Although the RT-size increases unremarkable, the misleading and traffic rise notably due to the summarization and collision.

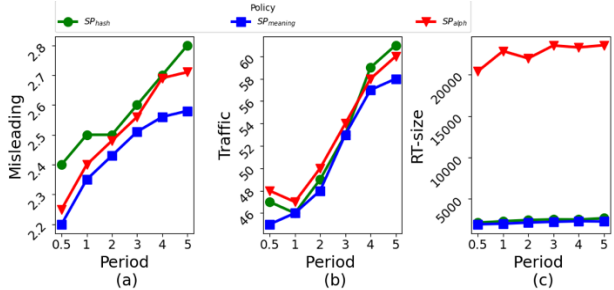

Fig. 22. *Periodically reestablishment threshold analysis*

### 9.3 Network Reestablishment Cost

This section studies the reestablishment cost for different network sizes from 1K to 30K nodes. The reestablishment cost measures the average number of messages needed for each advertised data stream. As shown in Fig. 23, the cost increases significantly when the network size increases from 1K to 15K. This is because the data in the routing table is still sparse, and the descriptors need to advertise all over the network. However, when the network size is larger than 15K, the cost slightly increases. This is understandable since the routing table indexed more data, and some data has existed. Summarization triggered also impact the advertisement cost. Therefore, our approaches can tolerate the network reestablishment for large size networks.

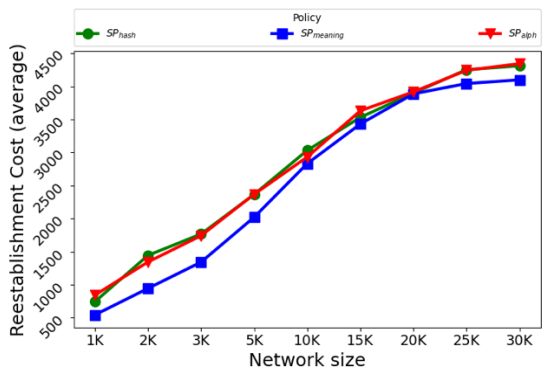

Fig. 23. *Network reestablishment cost analysis*

## 10 CONCLUSIONS

We explored the p2p data discovery problem in very large and growing IoT networks. Specifically, we delved deeply into the summarization techniques for MAA (and multi-keyword) based IoT data discovery. To achieve space efficiency for discovery routing, we not only analyzed potential summarization strategies, but also designed the novel coding scheme for each summarization strategy and explored the coverage issues. Our work is the first to make summarization for keyword based discovery routing truly feasible. For experimental studies, we crawled the web to collect a large dataset (posted on GitHub). Experimental results demonstrate the effectiveness of our summarization solutions and show that they outperform other p2p discovery solutions significantly.

**Hieu Tran** is currently an Ph.D. Candidate in Software Engineering at the University of Texas at Dallas, USA. He received the BSc degree in Software Engineering from Volgograd State Technical University, Russia, in 2016. His research interests include Program Analysis, Transfer Learning, IoT Data Discovery, and Cloud Computing. He is a member of the IEEE and the IEEE Computer Society.

**Son Nguyen** is currently an Ph.D. Candidate in Software Engineering at the University of Texas at Dallas, USA. He received the BSc degree in Computer Science from Vietnam National University, Hanoi, in 2015. His research interests include Service Composition and Automated Software Engineering.

**I-Ling Yen** received the BS degree from Tsing-Hua University, Taiwan, and the MS and PhD degrees in computer science from the University of Houston. She is currently a professor of computer science at the University of Texas at Dallas. Her research interests include parallel and distributed systems, cloud computing, service computing, security systems and algorithms, and high-assurance systems. She has been actively involved in several international conferences and workshops, serving as the General Chair, Organizing Committee member, and Program Committee member. She has served as a program chair/cochair for the IEEE Symposium on Service-Oriented Systems Engineering, IEEE Symposium on Reliable Distributed Systems, IEEE High Assurance Systems Engineering Symposium, IEEE International Computer Software and Applications Conference, IEEE International Symposium on Autonomous Decentralized Systems, etc. Currently, she is a steering committee member of the IEEE High Assurance Systems Engineering Symposium. She also serves on the editorial boards of the IEEE Transactions on Service Computing, the International Journal on Artificial Intelligence Tools, and the Knowledge and Information Systems journal.

**Farokh Bastani** received the BTech degree in electrical engineering from the Indian Institute of Technology, Bombay, in 1977, and the MS and PhD degrees in computer science from the University of California, Berkeley, in 1978 and 1980, respectively. He is the excellence in Education Chair professor of computer science at the University of Texas at Dallas and the director of the UTD site of the National Science Foundation (NSF) Net-Centric Software and Systems Industry/University Cooperative Research Center (NSF NCSS I/UCRC). His research interests include various aspects of systems engineering, especially the engineering of ultrahigh reliable software for safety-critical embedded systems, AI-based automated software synthesis and testing, embedded real-time process-control and telecommunications systems, formal methods and automated program transformation, high-assurance autonomous decentralized systems, high-confidence software reliability and safety assurance, inherently fault-tolerant and self-stabilizing distributed systems, modular parallel programs, and tele-collaborative systems. He was the editor-in-chief of the IEEE Transactions on Knowledge and Data Engineering (IEEE-TKDE) and has served on the editorial boards of the International Journal of Artificial Intelligence Tools, the International Journal of Knowledge and Information Systems, and the Springer-Verlag series on Knowledge and Information Management.